\documentclass[]{aa}  
\usepackage{soul}
\usepackage{graphicx}
\usepackage{multicol}
\usepackage{subfigure}
\usepackage{float}
\usepackage{ulem}
\usepackage{txfonts}
\usepackage[]{hyperref}
\hypersetup{colorlinks=true,citecolor=blue}
\begin{document}

   \title{Extended Lyman-$\alpha$ emission towards the SPT2349-56 protocluster at $z=4.3$}
   
   \author{Yordanka Apostolovski \inst{1}
            \and Manuel Aravena \inst{2}
            \and Timo Anguita \inst{3} \fnmsep \inst{4} 
            \and Matthieu Bethermin \inst{5}
            \and James Burgoyne \inst{6}
            \and Scott Chapman \inst{7}
            \and Carlos De Breuck \inst{8}
            \and Anthony Gonzalez \inst{9}
            \and Max Gronke \inst{10}
            \and Lucia Guaita \inst{3}  
            \and Yashar Hezaveh \inst{11} \fnmsep \inst{12}
            \and Ryley Hill \inst{7}
            \and Sreevani Jarugula \inst{13}
            \and Evelyn Johnston \inst{2}           
            \and Matt Malkan \inst{14}
            \and Desika Narayanan \inst{9}
            \and Cassie Reuter \inst{13}
            \and Manuel Solimano \inst{2}
            \and Justin Spilker \inst{15}
            \and Nikolaus Sulzenauer \inst{16}
            \and Joaquin Vieira \inst{13}
            \and David Vizgan \inst{13}
            \and Axel Wei\ss \inst{16}
}
   
   \institute{{Instituto de F\'{\i}sica y Astronom\'{\i}a, Universidad de Valpara\'{\i}so, Avda. Gran Bretaña 1111, Valpara\'{\i}so, Chile } \\
   \email{yordanka.apostolovski@gmail.com}
    \and {Instituto de Estudios Astrof\'{\i}sicos, Facultad de Ingenier\'ia y Ciencias, Universidad Diego Portales, Av. Ej\'ercito 441, Santiago, Chile}
    \and {Instituto de Astrofisica, Facultad de Ciencias Exactas, Universidad Andres Bello, Fernandez Concha 700, Santiago, Chile} 
    \and {Millennium Institute of Astrophysics, Monse\~{n}or Nuncio Sotero Sanz 100, Oficina 104, Santiago, Chile} 
    \and {Aix Marseille Univ, CNRS, LAM, Laboratoire d'Astrophysique de Marseille, Marseille, France}
    \and {Department of Physics and Astronomy,  University of British Columbia, Vancouver, BC,  Canada}
    \and {Department of Physics and Atmospheric Science, Dalhousie University, Halifax, NS, B3H 4R2, Canada}
    \and {European Southern Observatory, Karl Schwarzschild Stra\ss e 2, 85748 Garching bei M\"{u}nchen, Germany}
    \and{Department of Astronomy, University of Florida, 211 Bryant Space Sciences Center, Gainesville, FL, 32611, USA}
    \and {Max Planck Institut fur Astrophysik, Karl-Schwarzschild-Stra\ss e 1, D-85748 Garching bei M\"{u}nchen, Germany}
    \and {Département de Physique, Université de Montréal, Montreal, Quebec, H3T 1J4, Canada}
    \and {Center for Computational Astrophysics, Flatiron Institute, 162 Fifth Avenue, New York, NY, 10010, USA}
    \and {Department of Astronomy, University of Illinois at Urbana-Champaign, 1002 West Green St., Urbana, IL, 61801, USA}
    \and {Department of Physics and Astronomy, University of California, Los Angeles, CA, 90095-1547, USA}
    \and{Department of Physics and Astronomy and George P. and Cynthia Woods Mitchell Institute for Fundamental Physics and Astronomy, Texas A\&M University, 4242 TAMU, College Station, TX 77843-4242, USA}
    \and {Max-Planck-Institut f\"{u}r Radioastronomie, Auf dem H\"{u}gel 69, D-53121, Bonn, Germany}
    }

% \abstract{}{}{}{}{} 
% 5 {} token are mandatory
 
  \abstract
   {Deep spectroscopic surveys with the Atacama Large Millimeter/submillimeter Array (ALMA) have revealed that some of the brightest infrared  sources in the sky correspond to concentrations of dusty star-forming galaxies (DSFG) at high redshift. Among these, the SPT2349-56 protocluster system at $z=4.304$ is amongst the most extreme examples due to its high source density and integrated star formation rate.}
  {We conducted a deep Lyman-$\alpha$ line emission survey around SPT2349-56 using the Multi-Unit Spectroscopic Explorer (MUSE) at Very Large Telescope (VLT)  in order to characterize this uniquely dense environment.}
 %  {We present the detection and study of Ly-$\alpha $ emitters associated toward the proto-cluster SPT2349-56 at z=4.3. We establish the redshifted Ly-emission galaxies in the protocluster environment. This will allow us to measure the spatial distribution of most galaxies and their velocity dispersion with respect to the Dusty Star-Forming Galaxies (DSFG). We measure the individual properties of all galaxies in the protocluster based on the Ly-$\alpha$ line intensity, the CO and [C\textsc{ii}] line observations and optical/near-IR imaging. Also resolve spatially Ly-$\alpha$ emission from SPT2349-A DSFG, providing an unique test to the existence of inflowing/outflowing gas from this source, check for interactions with companions and inspect in its environment for possible extended nebulae.}
 %  {Through a tomography obtained with a deep 3d spectroscopic survey obtained with the Multi-Unit Spectroscopic Explorer (MUSE) located in the Very Large Telescope (VLT), perform a blind search inside the wavelength width of the proto-cluster to derive the redshifts of members galaxies. Identify within smooth narrow-band imaging, structures bounded to SPT2349-56 }
 {Taking advantage of the deep three-dimensional nature of this survey, we performed a sensitive search for Lyman-$\alpha$ emitters (LAEs) toward the core and northern extension of the protocluster, which correspond to the brightest infrared regions in this field. Using a smoothed narrowband image extracted from the MUSE datacube around the protocluster redshift, we searched for possible extended structures.}
 {We identify only three LAEs at $z=4.3$ in this field, in concordance with expectations for blank-fields, and an extended Lyman-$\alpha$ structure spatially associated with core of the protocluster. All the previously-identified DSFGs in this field are undetected in Lyman-$\alpha$ emission, consistent with the conspicuous dust obscuration in these systems. We find an extended Lyman-$\alpha$ structure, about $60\times60$ kpc$^2$ in size, and located 56 kpc west of the protocluster core. Three DSFGs coincide spatially with the location of this structure. We conclude that either the three co-spatial DSFGs or the protocluster core itself are feeding ionizing photons to the Lyman-$\alpha$ structure.}
 {}


   \maketitle
%
%________________________________________________________________

\section{Introduction}
\begin{table*}[tbh!]
\caption{Results of the blind line, narrow band and prior selected sample search in the MUSE data cubes.}              % title of Table
\label{tab:blind}      % is used to refer this table in the text
\centering
\begin{tabular}{ccccccc}
\hline\hline                        % inserts double horizontal lines
ID & RA & DEC & $\Delta v^\ddag$ & FWHM & $S_{\text{Ly}\alpha}$ & EW  \\    % table heading
  $\ldots$& (J2000) & (J2000) & (km s$^{-1}$) & (km s$^{-1}$) & ($10^{-20}$ erg cm$^{-2}$ s$^{-1}$) & (km s$^{-1}$) \\
\hline

LAB     & $23$:$49$:$43.39$ & $-56$:$38$:$23.49$   & $200 \pm 19 $ & $ 760 \pm 40$ & $3660 \pm 1030$ & $ 1530$\\ % LAB
LAE1	&	$23$:$49$:$41.28$ & $-56$:$37$:$58.20$ & $ -99 \pm 9	$ & $330 \pm 20$ & $ 510 \pm 150$ & $540$\\ % 21
LAE3	&	$23$:$49$:$42.18$ & $-56$:$38$:$10.48$ & $ 70 \pm 9	$ & $330 \pm 50$ & $ 2220 \pm 280$ & $1230$\\ % 72
LAE8    &   $23$:$49$:$40.03$ & $-56$:$37$:$34.11$ & $1689 \pm 9  $ & $330 \pm 30$ & $ 2300 \pm 430$ & $1300$\\
\hline
Tentative candidates$^\dagger$\\ 
\hline
NL3     &   $23$:$49$:$44.73$ & $-56$:$38$:$39.99$ & $-1818 \pm 39  $ & $270 \pm 90$ & $ 210 \pm 150$ & $1705$\\
LAE2	&	$23$:$49$:$44.26$ & $-56$:$38$:$40.90$ & $739 \pm 19	$ & $330 \pm 50$ & $ 190 \pm 170$ & $570$\\ % 56
LAE4	&	$23$:$49$:$39.90$ & $-56$:$38$:$12.22$ & $360 \pm 19	$ & $380 \pm 30$ & $ 240 \pm 170$ & $830$\\ % 109
LAE5	&	$23$:$49$:$43.47$ & $-56$:$37$:$02.37$ & $0 \pm 19	$ & $440 \pm 50$ & $ 370 \pm 240$ & $670$\\ %74
LAE6	&	$23$:$49$:$40.96$ & $-56$:$37$:$09.18$ & $669 \pm 19	$ & $310 \pm 50$ & $ 330 \pm 200$ & $560$\\ %91
LAE7	&	$23$:$49$:$45.41$ & $-56$:$37$:$28.70$ & $709 \pm 9	$ & $320 \pm 30$ & $ 320 \pm 220$ & $1170$\\ %118

\hline

\end{tabular}
\\
\noindent Notes: $^\dagger$ List of Lyman-$\alpha$ line candidates, which showed (snr)$_{\rm{det}}>5$ as computed in the LSDCat detection cube. Despite the high snr obtained in the ``maximal'' LSDCat extraction, these detections are considered tentative based on their low significance measured in the original cube through homogeneous $1\arcsec$ radii aperture measurements. $^\ddag$ Velocity offset relative to the protocluster's mean [C\textsc{ii}]-derived redshift, $z=4.304$.
\end{table*}

%LAB & $23$:$49$:$43.39$ & $-56$:$38$:$23.49$ & $366 \pm 16 $ & $ 760 \pm 37$ & $3663 \pm 1028$ & $ 1531$\\ % LAB
%LAE1	&	$23$:$49$:$41.28$ & $-56$:$37$:$58.20$ & $67	\pm 9	$ & $334 \pm 21$ & $ 514 \pm 146$ & $544$\\ % 21
%LAE3	&	$23$:$49$:$42.18$ & $-56$:$38$:$10.48$ & $242	\pm 5	$ & $326 \pm 45$ & $ 2216 \pm 283$ & $1230$\\ % 72
%LAE8    &   $23$:$49$:$40.03$ & $-56$:$37$:$34.11$ & $1863  \pm 12  $ & $334 \pm 28$ & $ 2301 \pm 434$ & $1297$\\
%\hline
%Tentative candidates$^\dagger$\\ 
%\hline
%NL3     &   $23$:$49$:$44.73$ & $-56$:$38$:$39.99$ & $-1646 \pm 38  $ & $266 \pm 89$ & $ 212 \pm 151$ &\\
%LAE2	&	$23$:$49$:$44.26$ & $-56$:$38$:$40.90$ & $906	\pm 22	$ & $326 \pm 53$ & $ 192 \pm 166$ & $566$\\ % 56
%LAE4	&	$23$:$49$:$39.90$ & $-56$:$38$:$12.22$ & $532	\pm 23	$ & $384 \pm 28$ & $ 240 \pm 171$ & $832$\\ % 109
%LAE5	&	$23$:$49$:$43.47$ & $-56$:$37$:$02.37$ & $170	\pm 22	$ & $441 \pm 51$ & $ 366 \pm 244$ & $673$\\ %74
%LAE6	&	$23$:$49$:$40.96$ & $-56$:$37$:$09.18$ & $836	\pm 19	$ & $307 \pm 45$ & $ 329 \pm 196$ & $556$\\ %91
%LAE7	&	$23$:$49$:$45.41$ & $-56$:$37$:$28.70$ & $880	\pm 12	$ & $324 \pm 28$ & $ 321 \pm 216$ & $1166$\\ %118

Studies of massive galaxies at the peak of their star-formation activity and their relation to the densest protocluster systems are key to understanding the hierarchical formation of the most massive galaxy structures in the early Universe. Current studies seek to understand the role of active galactic nuclei (AGN) feedback \citep{pike14, smolcic17}, or the relation between downsizing and star formation \citep{magliocchetti13, miller15} during the active growth phases of such forming structures. 

Cosmological simulations show that cold dark matter (CDM) haloes merge and form a web-like network traced by young galaxies and reionized gas. A protocluster will form at the highest overdensity regions within this filamentary structure at early cosmic times \citep[$z\sim4-6$;][]{baugh98,delucia07}, eventually becoming a massive virialized cluster by $z<1$ \citep[e.g.;][]{overzier16}. These cosmological simulations indicate that galaxies within galaxy protoclusters experience a luminous starburst-phase \citep{miley08}.
%As it show \cite{miley08} proto-clusters are expected to be luminous starbursts where massive galaxies and richest clusters come from regions with large overdensities. 
%If we compare cosmological simulations they present high over-densities of dark matter\citep{springel05,overzier09,chiang17} , so proto-clusters are expected to be luminous starbursts as show \cite{miley08}. However, observations show that the cluster-core formation is faster than predicted by simulations \citep{chapman09, chiang12, ma15}.

To identify and study these starbursting protocluster systems, several observational methods have been used. One of them corresponds to sub/millimeter wavelength observations, which allow one to pinpoint the obscured star-formation activity in young protocluster members \citep[e.g.,][]{chapman09, daddi09, aravena10, capak11, casey15, miller18, oteo18}. Similarly, low-frequency radio observations are typically used to search for radio-loud quasars sitting in the centers of dense protocluster fields  \citep{galametz13,rigby14}. In these radio-selected protoclusters, Lyman-$\alpha$ emitters (LAEs), star-forming galaxies selected through their significant UV rest-frame Lyman-$\alpha$ emission line ($\lambda_{\text{rest}}=1215.67 \AA$), show overdensity factors $3-5$ times larger than the field at the same redshift \citep{venemans05,venemans07}. Given this ubiquity of LAE overdensities in radio-selected fields, deep searches for these sources have been performed to confirm the redshifts of protocluster galaxy candidates using 8m-class optical/IR telescopes \citep[e.g.,][]{pentericci97, kurk00, venemans02, venemans04, venemans05, venemans07, croft05}.

The velocity dispersions of radio-selected galaxy protoclusters are typically found in the $\sim 300 - 1000$ km s$^{-1}$ range centered at the mean velocity of the radio galaxies. Although these systems are not yet virialized, such large velocity dispersions suggest that these systems have large halo masses, possibly evolving into the most massive cluster systems in the local Universe.

Narrowband image surveys in protocluster fields have identified a population of LAEs with luminosities larger than $10^{43.4}$ erg s$^{-1}$ and large spatial extensions ($40-150$ kpc). These structures are often referred to as Lyman-$\alpha$ blobs \citep[LABs;][]{steidel00,matsuda04}. The origin of the emission of these structures can be explained by different scenarios such as the presence of AGN or massive star-forming galaxies. The production of Lyman-$\alpha$ photons in these objects could be associated with different processes such as recombination radiation, continuum pumping, or collisional excitation \citep[see;][]{cantalupo17}.

The large-scale millimeter survey covering 2500 squares degrees of the sky conducted with the South Pole Telescope \citep[SPT;][]{carlstrom11} discovered a population of millimeter-bright sources  \citep[$S_{1.4\rm{mm}}>20$ mJy;][]{vieira10, vieira13, everett20, reuter20}. Follow-up observations with the Atacama Large Millimeter/submillimeter Array (ALMA) showed that the majority of these sources are gravitationally lensed submillimeter galaxies ($>90\%$; SMGs; also known as dusty star-forming galaxies, or DSFGs) with magnifications $\mu_{870\mu {\rm m}} \sim 5-20$ \citep[median $\mu_{870\mu {\rm m}} =6.3$][]{spilker16}. The remaining sources show no evidence of gravitational lensing, being either intrinsically bright or composed of fainter multiple-component SMGs or groups of SMGs \citep{hezaveh13,spilker16}. The high number of SMGs spread over a small area of the sky ($<1'$) found in these fields strongly suggests the existence of (sub)millimeter bright protocluster fields \citep{wang21}.

\begin{figure}
    \centering
    \includegraphics[width=1\linewidth]{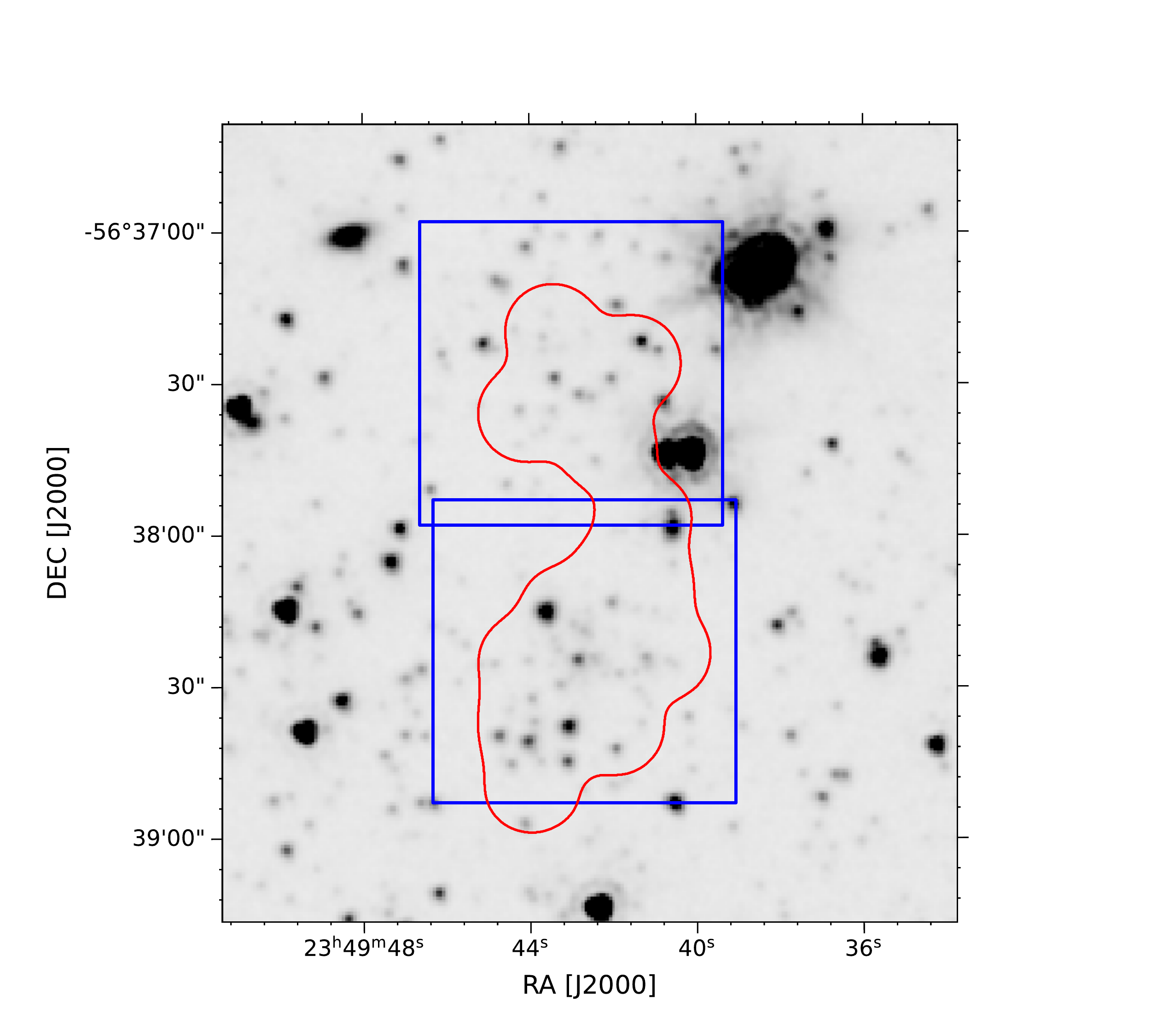}
    \caption{Deep IRAC mosaic obtained toward the SPT2349-56 system. Red contours show the ALMA [C\textsc{ii}] coverage. Blue squares show the observed MUSE footprint, where we used two pointings to cover the full IR bright region previously detected with LABOCA.}
    %LABOCA 870$\mu$m map toward the SPT2349-56 system. Red contours show the LABOCA emission, starting at 3$\sigma$, and increasing in steps of 3$\sigma$. Green and pink dashed circles represent the ALMA Band 3 and Band 7 pointings observations, respectively \citep{hill20}. Blue squares show the MUSE footprint, where we used two pointings to cover the full IR bright region, used in this paper.}
    \label{fig:laboca}
\end{figure}

Among the sample of SPT protoclusters, the SPT2349-56 system stands out due to its exceptionally high surface density of SMGs. SPT2349-56 is located at $z=4.304$ and has a surface density of more than ten times the average blank-field value and a volume density 1000 times the average \citep{miller18, hill20}. This system could represent the core of a massive galaxy cluster and is one of the most massive structures known to date in the early Universe. The SPT2349-56 system has two main infrared (IR) bright structures as seen in the APEX/LABOCA 870 $\mu$m maps, following the north-south direction (Figure \ref{fig:laboca}). The southern component comprises by itself a flux density of S$_{870} \approx 77$ mJy, whereas the northern component contributes with S$_{870} \approx 33$ mJy. For reference, a typical unlensed SMG has a flux density of around 5–10 mJy at 870\,$\mu$m. Higher-resolution deep ALMA spectroscopy yielded a total of 24 [C\textsc{ii}] and 16 CO(4-3) line emitters in the southern and northern extensions of the cluster \citep[e.g.][]{miller18,hill20}. Several components of this system have SFR estimates of $\sim1000$ M$_\odot$ yr$^{-1}$, while the full protocluster system is estimated to have a SFR of about $6.6 \times 10^{4}$ M$_\sun$  yr$^{-1}$ \citep{hill20}. Similarly, the dynamical mass of the core region is estimated to be $\sim9\times10^{12}\ M_\sun$, while the total halo mass of the whole structure is $\sim2.5\times10^{13}\ M_\sun$ \citep{hill20}.

The physical properties of these sources indicate that this protocluster already harbors massive galaxies that are rapidly forming stars from an abundant gas supply. The large number of SMGs in this system pushes and challenges theoretical models seeking to explain the origin and evolution of protoclusters \citep{chiang12}. 

Due to the proximity of the SMG members in the core of the protocluster (the diameter is about $130$ kpc), it is likely that its component galaxies will merge to form a massive elliptical galaxy at the core of a lower-redshift Coma-like galaxy cluster \citep{miller18, hill20}.

A recent search for Lyman Break Galaxies (LBGs) in the extended SPT2349-56 environment found 4 LBGs in the southern part of the protocluster \citep{rotermund21}, indicating that most of the SMGs are inconspicuous at optical wavelengths, with only one of the 4 LBGs coinciding with a previously reported SMG. % PUEDE CAMBIAR DE LUGAR 

Motivated by the significant overdensities found in radio-selected protocluster fields, we conducted an independent census of star-forming galaxies in the SPT2349-56 field through a sensitive search for Lyman-$\alpha$ emission using deep optical spectroscopy obtained with the Multi-Object Spectroscopy Unit (MUSE) at the Very Large Telescope (VLT). In Section \ref{sec:obs} we describe the observations and reduction of the MUSE data towards SPT2349-56, and summarize previous observations.  In Section \ref{sec:res} we present the detection of Lyman-$\alpha$ emission through both a blind search and narrowband imaging. In Section \ref{sec:ana} we analyze the nature of the extended Lyman-$\alpha$ emission along with its connection with LAEs and the structure of the protocluster. Section \ref{sec:summ} summarizes and presents the conclusions of this work.

Hereafter, we adopt a flat $\Lambda$CDM cosmology with $h=0.677$, $\Omega_m=0.307$ and $\Omega_\Lambda=0.693$ \citep{planck15}.

\section{Observations} \label{sec:obs}

In this section, we describe details of the Lyman-$\alpha$ line observations in the SPT2349-56 protocluster at $z=4.3$. 

\begin{figure*}[ht]
\centering
\includegraphics[width=0.85\textwidth, trim={3cm 0 3cm 0}]{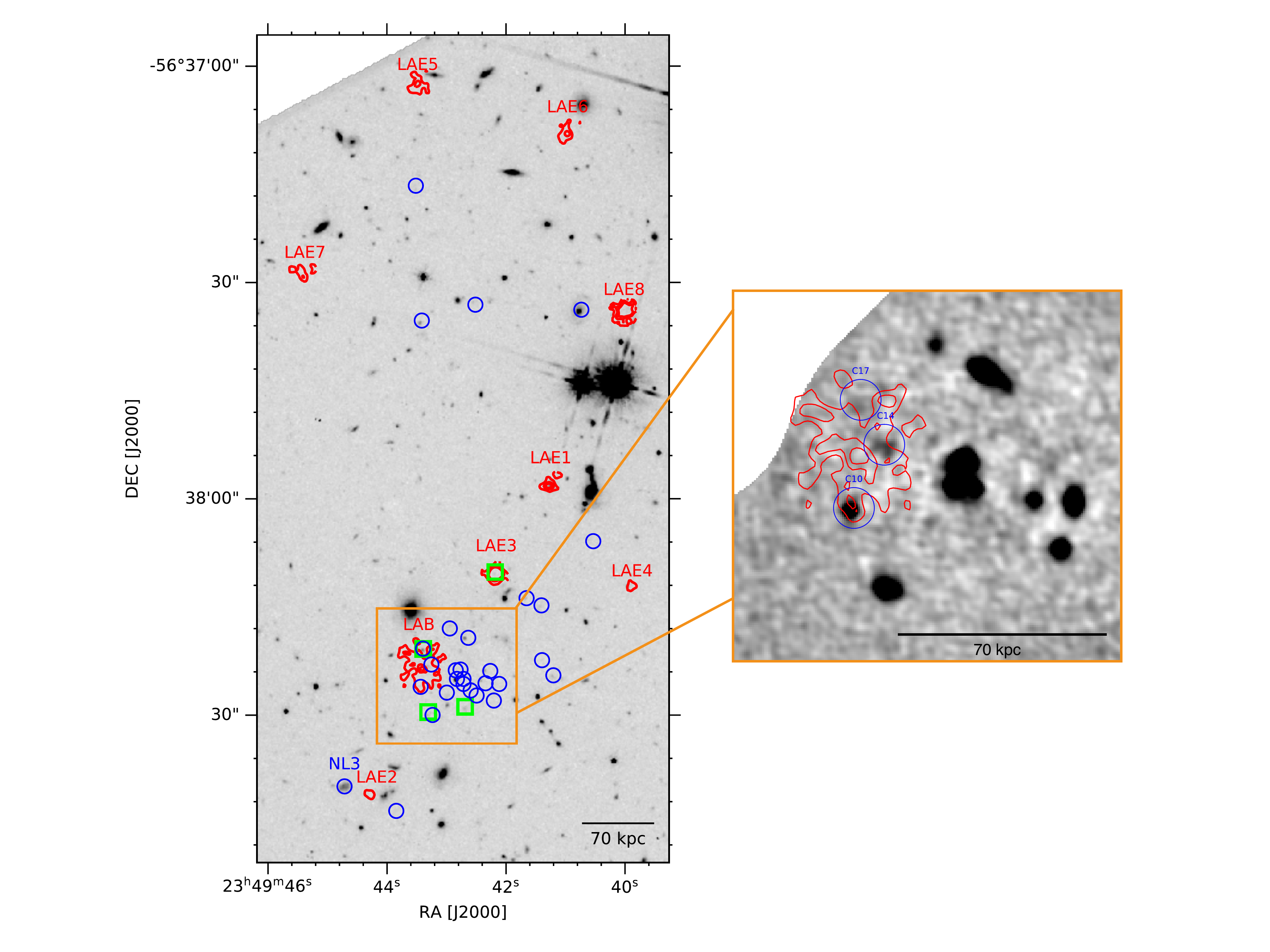}
\caption{\textit{Left:} Lyman-$\alpha$ emission toward the SPT2349-56 protocluster system at $z=4.3$. The MUSE covered area is shown, with the HST F160W image in the background in grey-scale, and red contours representing a rendered Lyman-$\alpha$ image. The later is obtained as the average of each individual line map of the detected LAEs and the LAB, in steps of 2, 5 and 7 $\sigma$, where $\sigma$ is the rms noise level in the average image. The blue circles highlight the location of the ALMA [C\textsc{ii}] and CO(4-3) line detections in the field \citep{miller18, hill20}. Green squares show the location of the LBGs in the field \citep{rotermund21}. \textit{Right:} The map of ALMA [C\textsc{ii}] line emission toward the identified Lyman-$\alpha$ blob (LAB) is shown in the background, with blue circles representing the location of the previously identified [C\textsc{ii}] line emitters C10, C14, C17 \citep[see Table 2;][]{hill20}. Red contours show the Lyman-$\alpha$ emission of the LAB at 2, 4, 6 and 8$\sigma$.}
\label{fig:hst}
\end{figure*} 

\begin{figure}
\centering
\includegraphics[width=1\linewidth,  trim={5cm 0 5cm 0}]{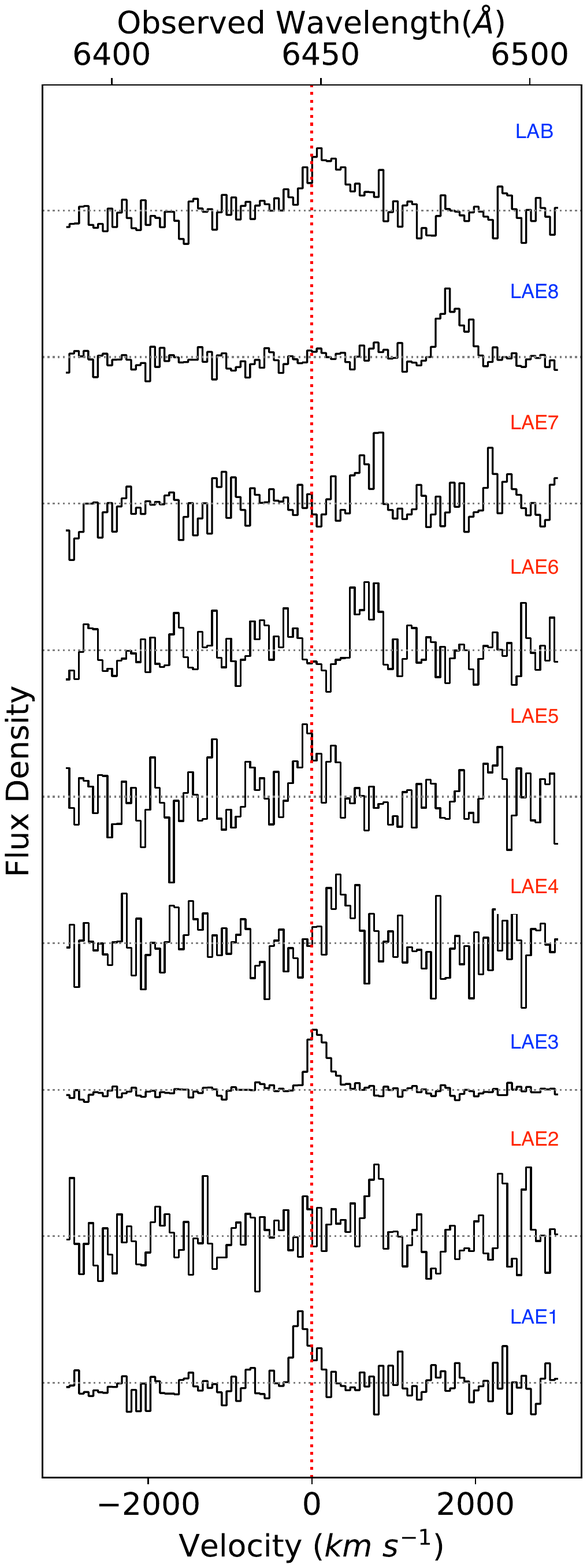}
\caption{Continuum subtracted spectra of the Lyman-$\alpha$ emission line identified significantly within the MUSE footprint around SPT2359-56. For comparison purposes, the vertical axis has been normalized. The measured fluxes are given in Table \ref{tab:blind}. Red dotted line shows the central velocity of the protocluster (expected redshifted Lyman-$\alpha$ line at $\lambda=6444.2 \AA$ ). The nomenclature of the LAEs does not follow the snr of the emission lines. The blue tag name denotes secure detections while the red tag names denote tentative detections.}
\label{fig:spectrum}
\end{figure}

\subsection{MUSE observations}

Observations with MUSE at the VLT UT4 were performed in two separate pointings targeting the north and south extensions of the SPT2349-56 protocluster system (Figure \ref{fig:laboca}). MUSE covers the wavelength range 480-930 nm. Each pointing covers roughly 1 square arcmin ($60''\times60''$). These observations were carried out in the wide field mode (WFM) in service-mode observing as a part of projects 0100.A-0437(A) and 0100.A-0437(B) (PI: M. Aravena) during dark-time. Each pointing was observed for 5 hours (a total of 10 hours) between November 2017 and September 2018. Each pointing consisted of a set of exposures of 680 seconds each, with individual exposures rotated by 90 degrees with respect to each other.

The average seeing of these observations was 0.97 and 0.98 arcsec for the southern and northern pointings, respectively, after correction of air-mass. Weather conditions were classified by ESO as clear (CL; 55\%), with high wind (CL-WI; 11\%) and photometric conditions (PH; 33\%) for all observing blocks (OBs).

We reduced the data using the MUSE pipeline v2.6 \citep{weilbacher14} for bias subtraction,  flat-fielding, and wavelength and flux calibration, resulting in a single data cube per each of the 5 OBs per field. We combined the five OB data cubes per field using the MUSE Python Data Analysis Framework (MPDAF; \citealt{mpdaf}). The data cubes were merged using a sigma-clipped mean with $\sigma_{\text{clip}}=5$. Since the field is relatively sparse (especially in Lyman-$\alpha$ at $z=4.304$), we used Zurich Atmosphere Purge (ZAP; \citealt{zap}) to perform a sky subtraction  through principal component analysis (PCA). For this process, we used a mask in order to avoid spaxels that contained obvious continuum sources.

\subsection{Previous ALMA observations}

In this study, we used as reference the images, cubes and location of protocluster members previously identified through ALMA Cycle 5 and 6 observations. These observations and the corresponding data reduction and source identification are described in detail by \cite{miller18} and \cite{hill20} and we refer the reader to those papers for full details. 

In brief, observations of the redshifted [C\textsc{ii}]$_{158\mu\text{m}}$ fine structure line towards the SPT2349-56 system were obtained using ALMA in Band 7. These were centered at a frequency of $\nu_{obs}=358.4$ GHz, yielding an average synthesized beam size of $0.43'' \times 0.34''$ and 3$\sigma$ sensitivities of $\approx 0.3$ mJy beam$^{-1}$. These observations, which cover the full IR-bright region, led to the identification of 24 [C\textsc{ii}] emitters in the field. The MUSE observations described above fully cover the region observed by ALMA in Band 7 at with uniform sensitivity (Fig. \ref{fig:laboca}). Based on the identified [C\textsc{ii}] sources, the mean redshift of the system was determined to be at $z=4.304$ \citep{miller18, hill20}.

\subsection{HST imaging}
We used HST/\textsl{Wide Field Camera 3 (WFC3)-IR} images under program 15701 (PI: S. Chapman). The target was assigned 2 orbits for the F110W filter and three orbits for the F160W filter in the infrared channels. Dithering was implemented for maximum resolution. The data was reduced using the standard \textit{HST} pipeline. The pixel size in the WFC3 images is $0.075''$.

\section{Results} \label{sec:res}

We used the MUSE observations obtained toward SPT2349-56 to perform a systematic search of Lyman-$\alpha$ emission with three methods: a blind automatic search in the cube, creating a narrow band image around the known protocluster redshift and a search for Lyman-$\alpha$ emission in (dusty) sources that had previously been identified in this field. Below, we describe each of these searches.

\subsection{Blind Search}

We performed a blind search for Lyman-$\alpha$ emission in the MUSE data cubes, using the Line Source Detection and Cataloguing Tool (\texttt{LSDCat}; \citealt{lsdcat}). For this, we focused on a 4000 km s$^{-1}$ band centered on the red-shifted ($z=4.304$) Lyman-$\alpha$ wavelength ($\lambda_{\text{red}} = 6444.2\  \AA$). The \texttt{LSDCat} routine detects emission lines through an spatial and spectral filtering (3D matched-filtering) approach and sorts them into discrete objects. This method is used to maximise the signal-to-noise ratio (snr) of the entire cube, and thus creating a snr detection cube. To determine an appropriate threshold for detection in the snr cube, we conducted an unbiased line search also in the negative version of our original cube (multiplied by $-1$). Assuming that the noise in this reduced velocity range of the cube is symmetric around 0 and roughly follows a Gaussian distribution, the detections obtained in the negative cube will set the maximum level at which we expect line features produced by noise. From this, we find that the most significant feature in the negative cube is found at (snr)$_{\rm{det}}$ $\sim 5$, thus yielding our detection threshold. 

This process yields a significant number of positive features located at the edges of the independent channel images, and with linewidths of one or two channels only, which we remove from our catalogue as they are unphysically narrow. To filter the Lyman-$\alpha$ line candidates from spurious positive features, we constrain the full-width at half maximum (FWHM) of the detected lines to the range of widths found for the [C\textsc{ii}] and CO(4-3) lines for sources in the field \citep{hill20}, which correspond to $50-600$ km s$^{-1}$. After this selection process, we identified eight LAE candidates, four in the northern and four in the southern pointing (Figure \ref{fig:hst}).

We extracted the spectra of each of the LAE candidates using apertures with radii of 1 arcsec. Due to the more extended spatial nature compared to the other LAEs, we extracted the spectra of sources LAE3 and LAE8 using apertures of 2 arcsec radius (Figure \ref{fig:spectrum}). Based on the significance of each of the line candidates measured in these apertures (see Fig. \ref{fig:spectrum} and Table \ref{tab:blind}), we split the sample in secured LAEs and tentative candidate sources. %This process further filters sources that were selected based on the (snr)$_\rm{det}$ threshold. 

Only three sources are securely detected in this fashion (LAE1, LAE3 and LAE8), and the remaining five sources are thus considered tentative detections. All the spectra were manually inspected and searched for other lines that would point to a lower redshift possibility. However, all line detections were consistent with Lyman-$\alpha$ at $z\sim4.3$. 

LAE1 is associated with detections in all available broadband images, including $g$, $r$, $i$ through $K_S$ band \citep[see Appendix B in][]{hill22}. However, if the galaxy is at $z\sim4.3$, we would expect it to be faint in the $g$ band due to the Lyman break. Inspection of the HST F110W image (see Fig. \ref{fig:cutouts}) suggest that the excess g-band emission comes from a foreground object along the line of sight. Indeed, due to the mismatch between MUSE Lyman-$\alpha$ position and the optical broadband position, \citet{hill22} lists its photometry as upper limits (see their Table 1). The significance of the detected line and the lack of other line features in the MUSE spectrum strongly favour the $z\sim4.3$ spectroscopic confirmation. As a precedent, note that ALMA source C1 (source `A') appears to be well detected in $g$-band, since there is a foreground $z=2.5$ galaxy as shown in \citet{rotermund21}.

LAE3 and LAE8 are both undetected in the $g$-band and have faint detections in the deep $r$ and HST F110W images \citep[see Fig. \ref{fig:cutouts} and Appendix B in][]{hill22}. They are significantly detected in Lyman-$\alpha$ without other line identifications, and have considerable EWs compared to the other sources identified in this field.

%Despite LAE1 being well detected in line emission, it shows continuum emission in \textit{g-band}, which would imply that the line identification is not Lyman-$\alpha$. However, we do not observe any additional line features in the full MUSE spectrum that would denote a different redshift in this galaxy, being thus more likely that our Lyman-$\alpha$ line identification is correct. 

Based on the redshift implied by the identified Lyman-$\alpha$ lines, we computed a median velocity offset for all the LAEs in the field with respect to the protocluster redshift $z=4.304$ \citep[obtained from previous \text{[C\textsc{ii}]} and CO identifications;][]{miller18, hill20}. The northern and southern LAEs are found to have velocity offsets of $\Delta v = 930$ and $\Delta v = 430$  km s$^{-1}$ respectively, indicating that the Lyman-$\alpha$ emissions are systematically redshifted from the center of the protocluster.
%\textcolor{red}{However, due the low snr in Lyman-$\alpha$ (Table \ref{tab:blind}) we tag five of the LAE as tentative candidates.}

\begin{figure*}[ht]
    \centering
    \includegraphics[width=1\linewidth]{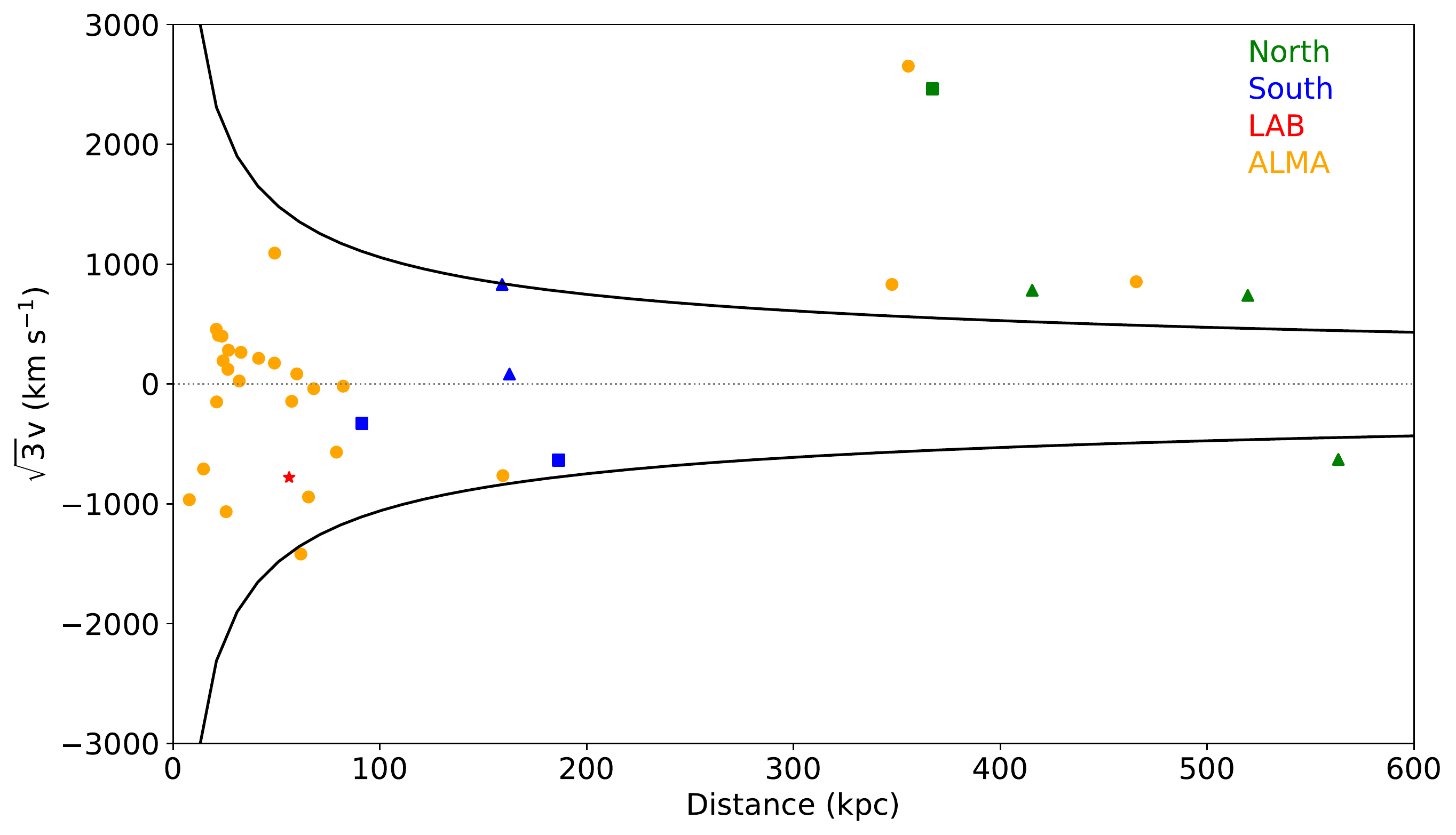}
    \caption{Velocity offset scaled by $\sqrt{3}$ as an estimate for 3-dimensional velocity, centered at the center of the protocluster \citep{hill20} versus projected distance from the 850 $\mu$m-weighted centre of the protocluster. Orange circles show detections of [C\textsc{ii}] and CO(4-3) with ALMA \citep{miller18,hill20}, green square and triangles show Lyman-$\alpha$ emitters detected and candidates respectively in the extended emission showed by LABOCA observations (Figure \ref{fig:laboca}), blue squares and triangles show the Lyman-$\alpha$ emitters detected and candidates respectively in the southern pointing and the red star shows the Lyman-$\alpha$ blob inside the 90 kpc defined as the effective radius. Black lines show the escape velocity from the protocluster. The measured velocity offset uncertainties are negligible, and thus errorbars are smaller than the size of the symbols.}
    \label{fig:bound}
\end{figure*}

\subsection{Narrowband Image}

To independently search for line emission in the field, we produced a continuum-subtracted narrowband image using a spectrally and spatially smoothed version of the MUSE datacube with LSDCat. We selected as a central wavelength for the image of the Lyman-$\alpha$ line redshifted to $z=4.304$ and a width of 2000 km s$^{-1}$ (i.e. $6401.3 - 6487.2 \AA$). As such, this procedure was specifically designed to search for extended emission. As a result, we found an extended Lyman-$\alpha$ structure towards the east of the protocluster core, which we associate with a so-called ``Lyman-$\alpha$ blob’’ (LAB, see Figure \ref{fig:hst}). The Lyman-$\alpha$ emission of the LAB subtends a roughly circular region with an area of $10''\times10.4''$ in the sky ($\approx$ 70 $\times$70), and is located about $56$ kpc east of the center of SPT2349-56. With a radius of $\approx5''$ (34.4 kpc), this yields an area, $\pi r^2=3720$ kpc$^2$ or $\sim60\times60$ kpc$^2$.

To obtain a spectrum of this extended emission, we draw a polygon around this source, containing all pixels detected above 2$\sigma$ in the narrowband image (see Figures \ref{fig:hst} and \ref{fig:spectrum}). Based on the Lyman-$\alpha$ profile, we find that the extended feature shows a line  with a FWHM of about $760$ km s$^{-1}$ and a velocity offset with respect to the [C\textsc{ii}]/CO protocluster redshift of $\Delta v = 365$ km s$^{-1}$. After integrating along the full width of the line emission we obtain a flux of $S_{\text{Ly}\alpha} = 3663 \times 10^{-20}$ erg cm$^{-2}$ s$^{-1}$. 

A comparison of the position of the LAB with the location of the previously-identified SMGs in this field \citep{miller18,hill20} shows that three sources  overlap spatially with the western part of the blob. These sources, called C10, C14 and C17 using the nomenclature of \cite{hill20}, were identified based on their bright [C\textsc{ii}] line emission, with fluxes of  2.96, 1.70 and 0.93 Jy km s$^{-1}$, respectively. These galaxies are not the most luminous in the sample of [C\textsc{ii}] emitters in the SPT2349-56 system, and are located at about 65 kpc from the center of the protocluster.  

\subsection{Previously Known Protocluster Members} \label{ALMAcounterparts}

In addition to the independent searches described above, we searched the MUSE datacubes for Lyman-$\alpha$ emission at the positions of the previously-identified SMGs in the SPT2349-56 system at $z=4.3$. All of these sources have confirmed systemic redshifts based on the identification of the [C\textsc{ii}] and CO(4-3) lines with ALMA \citep{hill20}.

For each of these sources, we extracted a spectrum using apertures with radii of $2\arcsec$ centered at the location of either the [C\textsc{ii}] and CO(4-3) detections (Figure \ref{fig:ALMA_detections}). Inside the range of 6000 km s$^{-1}$ centered at $z=4.304$ we do not find significant Lyman-$\alpha$ emission in any of the previously-confirmed SMGs in the field. However, we do find a tentative detection of Lyman-$\alpha$ emission from one of the ALMA continuum sources in this field for which no redshift confirmation was possible using the [C\textsc{ii}] or CO(4-3) lines, source NL3. This source shows possible Lyman-$\alpha$ emission at a velocity of -1600 km s$^{-1}$ from the cluster core redshift ($z=4.304$). This velocity is covered by the ALMA CO observations but not by the [C\textsc{ii}] ones. Thus, while the ALMA [C\textsc{ii}] observations missed the line, it is possible that either the source is too faint in CO(4-3) emission or the tentative Lyman-$\alpha$ feature is not real. We thus tag this as a tentative candidate here.

We note that stacking of the MUSE spectra to yield a constrain on the average Lyman-$\alpha$ emission in these undetected SMGs is difficult. Several studies have demonstrated that the Lyman-$\alpha$ emission line does not always trace the galaxies' systemic redshifts due to IGM scattering and absorption \citep[e.g.][]{shapley03, song14, hashimoto15}. Therefore, aligning the MUSE spectra at the [C\textsc{ii}] or CO-derived systemic redshifts or correcting them to the rest-frame will yield a diluted Lyman-$\alpha$ stack signal. While corrections for the Lyman-$\alpha$-derived to the systemic redshifts (or velocities) as a function of equivalent width have been calibrated, these are statistical in nature and will not yield the precise redshift/velocities necessary for stacking. We do use such corrections in the check to see if the identified LAEs are gravitationally bound to the protocluster core (see next section).

\section{Analysis and discussion} \label{sec:ana}

\begin{figure}[t!]
    \centering
    \includegraphics[width=1\linewidth]{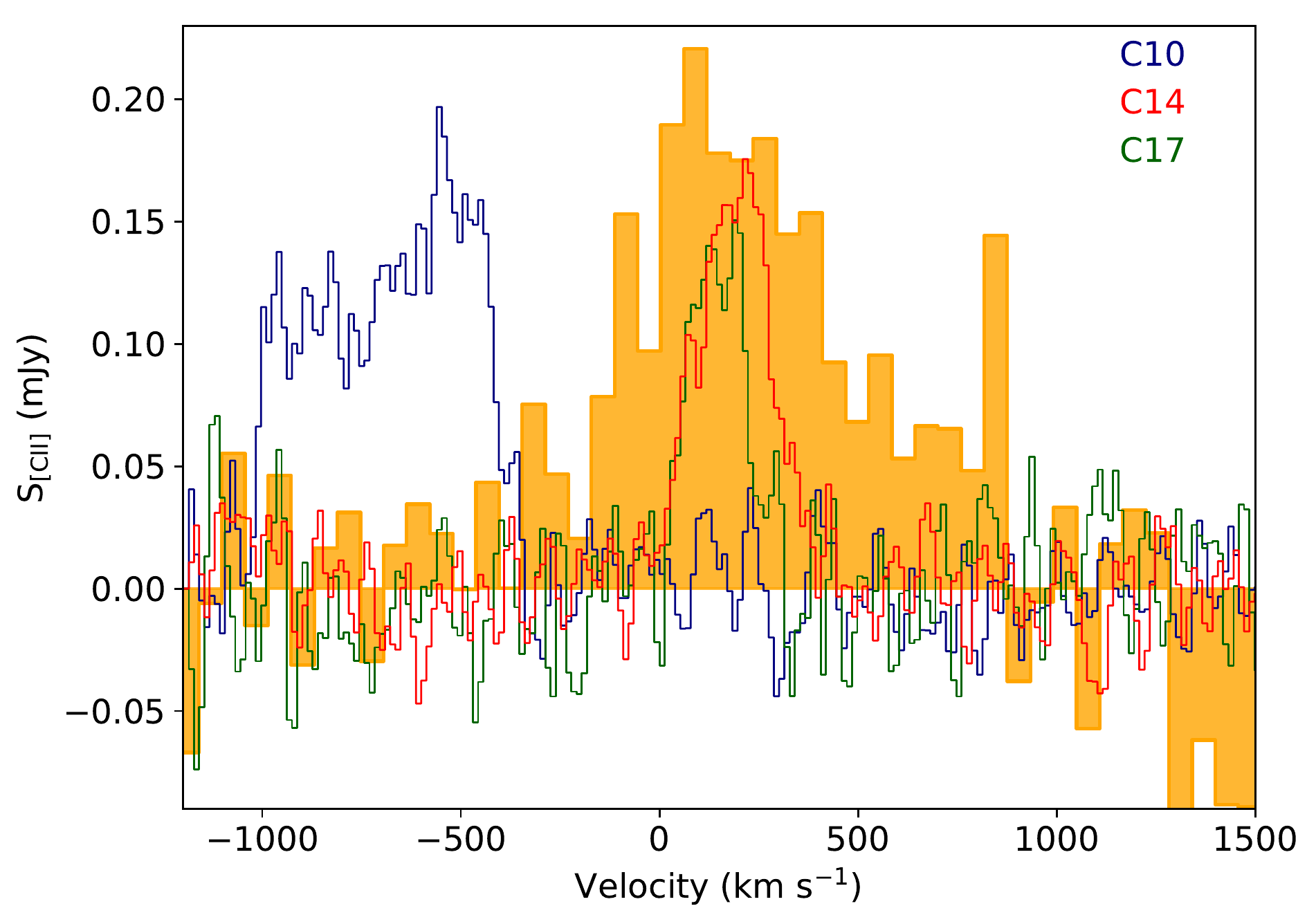}
    \caption{Observed [C\textsc{ii}] line profiles obtained with ALMA for sources C10, C14 and C17 \citep[see,][]{hill20}, which overlap spatially with the Lyman-$\alpha$ Blob obtained with MUSE (see Fig.\ref{fig:hst}). The scaled Lyman-$\alpha$ profile is shown in orange, for reference. The velocity scale refers to $z=4.304$ for all emission lines.}
    \label{fig:blobcii}
\end{figure}

\subsection{Lyman-$\alpha$ emitters}

As we showed in the results section, we have found 8 blindly selected LAEs and one LAB. Three of the blind identifications are considered secure and 5 are tentative based on their low significance (see Table \ref{tab:blind}). As shown in Fig. \ref{fig:hst}, the LAB and four of the identified LAEs are located in the southern structure, while the rest are located in the north. The locations of the LAEs and candidates range from 95 to 580 kpc from the core of the protocluster and the implied velocities for Lyman-$\alpha$ appear redshifted with respect to its systemic redshift. 

The significant mass of the central core of the protocluster is $\sim 9 \times 10^{12} M_\odot$, making it possible that most of the galaxies identified in the core neighbourhood are gravitationally bound as already shown by \cite{hill20}. To test whether the identified Lyman-$\alpha$ emitters are also bound, which may yield insights into the future evolution of the LAEs within the SPT2349-56 system, we compare the offset velocities of each of the protocluster members as a function of distance from the protocluster core (Fig. \ref{fig:bound}). For the ALMA-identified protocluster members we show their [C\textsc{ii}]/CO-based velocities, while for the LAEs, we show their Lyman-$\alpha$-based velocities. For the later, we obtain a statistical correction to the systemic redshift of each galaxy using the relation described in \citet{verhamme18}. Galaxies that have velocities lower than the the escape velocity envelope at a given radius from the protocluster core are expected to be bound to the structure. The southern LAEs appear to be consistently associated to the protocluster core, including the LAB. Only one of the identified LAEs in the southern structure (a tentative candidate) resides outside the escape velocity $v_\text{esc}=\sqrt{2GM/R}$ envelope. Conversely, most of the northern LAEs, including the secure LAE8 identification, appear to be unbound and redshifted with respect to the protocluster core yet following the trend of the other [C\textsc{ii}] identified sources. The fact that the northern LAEs follow the velocity offset of the ALMA [C\textsc{ii}] sources in the north, bolsters the interpretation of the northern structure as an unbound/infalling sub-halo \citep[e.g.,][]{miller18,hill20}. In addition, this supports the idea that the southern structure, which includes the protocluster core is likely already reaching a virialized form.

%Three out of the four LAEs in the northern structure appear to be unbound, whereas most of the southern LAEs appear to be consistently associated to the protocluster core, including the LAB. This supports the idea that the southern structure, which includes the protocluster core is likely already reaching a virialized form.

We compare the number of Lyman-$\alpha$ emitters found in the SPT2349-56 field at $z=4.3$ with the field counts using the ultra-deep MUSE observations in the Hubble Ultra Deep Field \citep[HUDF \textit{mosaic} area of $3\times3$ arcmin$^2$]{inami17}. Down to a Lyman-$\alpha$ luminosity of log($L_{{\rm Ly}\alpha})=41.0$, corresponding to the depth reached by our SPT2349-56 observations, \citet{drake17} finds 144 LAEs in the redshift range $z=4.0-5.0$. The MUSE HUDF observations are $>2\times$ deeper than our pointings, and thus are complete to this depth. Considering the redshift range used to search for LAEs in the SPT2349-56 ($z=4.25-4.36$) and the MUSE covered area ($2$ arcmin$^2$), we would thus expect to have found $3-4$ LAEs. This indicates that the detection of 3 secure sources in the SPT2349-56 field are consistent with blank-field counts, and strengthens the case that most of the emission output and mass in this system is associated to heavily dust obscured sources. In this scenario, the existence of a LAB in such complex obscured environment appears as a rare case, where the Lyman-$\alpha$ emission is able to escape in a preferential, less obscured direction.

%Since in reality we find 9 LAEs in the SPT2349-56 field (including the LAB), this yields an overdensity of $\sim2-3$ with respect to the HUDF. This is significantly lower than the overdensity of SMGs found in this field \citep{miller18}. 

\begin{figure}[ht]
    \includegraphics[width=1\linewidth]{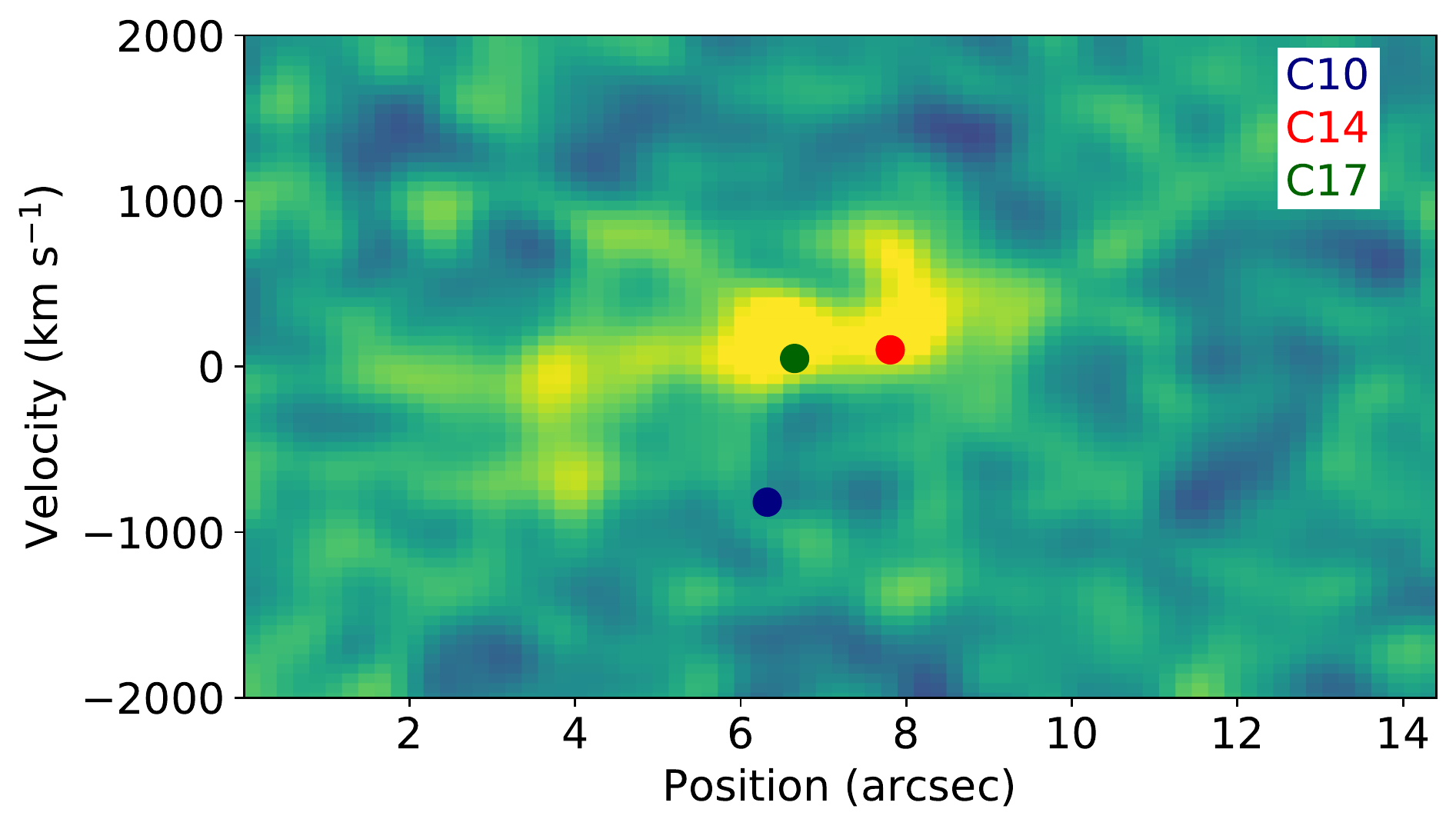}
    \caption{Position-velocity diagram towards the Lyman-$\alpha$ Blob, extracted from MUSE cube at 0 degrees of inclination towards the east. Blue, red and green circles show the observed [C\textsc{ii}] emissions (C10, C14 and C17) from \cite{hill20}. This shows a spatial connection between the LAB and two DSFGs members of the PC.}
    \label{fig:pv}
\end{figure}

\subsection{Insights on the nature of the extended Lyman-$\alpha$ emission}

As mentioned previously, the extended Lyman-$\alpha$ emission found in the MUSE-based narrowband image broadly coincides with the position of three SMGs that were identified as part of the protocluster structure. This suggests a possible physical relationship between them as indicated by previous studies of LABs at high redshift \citep[e.g., ][]{chapman01,umehata15,geach16,oteo18}. 
 
\cite{cen13} constructed a model for the origin of extended Lyman-$\alpha$ emission in the context of the cold dark matter framework, in which the LAB are produced due to starburst activity. The model incorporates AGN feedback, although it is expected that it should have a subdominant contribution \citep[e.g.,][]{webb09}. For extended Lyman-$\alpha$ emission in a protocluster, each galaxy member contributes to the whole Lyman-$\alpha$ emission yielding a variety of sizes and geometries typically found within a contiguous structure. The relative contribution of these DSFGs depends on the dust attenuation of Lyman-$\alpha$ photons and the propagation and diffusion process through the circumgalactic medium (CGM) and intergalactic medium (IGM) of each member.
 
%In the context of the cold dark matter model, \cite{cen13} developed a starburst model for the origin of LABs, with and without central AGN feedback. This, due to the fact that a large fraction of the stellar and AGN optical to UV radiation is absorbed by dust and reemitted less energetic wavelengths, being the AGN contribution the subdominant mechanism\citep[e.g.,][]{webb09}. For extended Lyman-$\alpha$ emission in a protocluster, each galaxy member contributes to the whole Lyman-$\alpha$ emission yielding a variety of sizes and geometries typically found within a contiguous structure. The relative contribution of these DSFGs depends on the dust attenuation of Lyman-$\alpha$ photons and the propagation and diffusion process through the circumgalactic medium (CGM) and intergalactic medium (IGM) of each member.

In Figure \ref{fig:blobcii} we compare the Lyman-$\alpha$ spectra of the LAB with the [C\textsc{ii}] emission line spectra of the three DSFGs spatially coincident to it: C10, C14 and C17. The [C\textsc{ii}] line emission is expected to trace the kinematics of each host galaxy, and thus probe the galaxies' systemic velocities and geometries (rotation, merger, etc). Due to obscuration and absorption by the intergalactic medium, the Lyman-$\alpha$ spectrum is expected to be redshifted with respect to the galaxies' systemic velocities, and thus with respect to the [C\textsc{ii}] lines. In this case, the Lyman-$\alpha$ spectrum of the LAB is found $\sim300$ km s$^{-1}$ redward of the protocluster core velocity ($v=0$ km/s) and $\sim100$ km s$^{-1}$ of the C14 and C17 SMGs. A much larger velocity difference is seen between the LAB and the systemic velocity of the C10 galaxy. The inconsistency in velocities for C10 suggest this source might be unrelated to the Lyman-$\alpha$ emission. We explore this issue in more detail in the next sections.

%Furthermore, we find that each of these three galaxies show narrow [C\textsc{ii}] line widths (median FWHM $\sim 320$ km s$^{-1}$) compared with that of the Lyman-$\alpha$ emission in the blob (FWHM $\sim 760$ km s$^{-1}$). However, C10 presents a wider, double peaked [C\textsc{ii}] line with a total width of $\sim 520$ km s$^{-1}$, which is just slightly narrower than the Lyman-$\alpha$ width of the blob. Along with the consistency in velocities for C14, C17 and the protocluster core, this suggest that the Lyman-$\alpha$ emission in the LAB is not associated with the kinematics of the individual galaxies alone, where the different lines would be expected to have similar linewidths, but instead associated to additional physical mechanisms. 

\subsection{The protocluster core as the origin for the LAB?}

The LAB is located $\sim 56$ kpc to the east of the of the protocluster core, thus being within the protocluster effective radius defined by \cite{hill20}. Along with the velocity connection between the LAB and the SMGs, this spatial coincidence suggest a physical link between the protocluster core and the LAB. It is thus possible that the powering source of the extended Lyman-$\alpha$ emission is star formation or AGN activity in the starbursting SMGs at the protocluster center, where the Lyman-$\alpha$ photons are produced in a photon-ionized medium.

%where \cite{hill20} predicts that the core could merge into a brightest cluster galaxy (BCG),

%Due to the dust obscuration within the protocluster core,  it is possible that Lyman-$\alpha$ emission exists near the protocluster core. 

In this scenario, it is possible that most of the Lyman-$\alpha$ photons along our line of sight are not absorbed and/or scattered but are instead able to escape toward the eastern part of the protocluster core. Indeed, \cite{vernet17} observed similar regions with offsets of $\sim 100$ kpc in the haloes of high redshift AGN-host galaxies, invoking similar arguments.

Following \cite{furlanetto05}, the Lyman-$\alpha$ emission can be used to yield an estimate of the underlying SFR from the powering source. For star formation episodes following a Salpeter initial mass function \citep{salpeter59} and that two thirds of the ionizing photons are absorbed in the dense ISM, we have:

\begin{equation}
    \centering
    L_{\text{Ly}\alpha}=10^{42} (\text{ SFR}/[M_\odot \ \text{yr}^{-1}]) \ \text{erg s}^{-1} 
    \label{eq:1}
\end{equation}

Taking the value of $L_{\text{Ly}\alpha} = 1.32 \pm 0.37 \times  10^{42}$ erg s$^{-1}$, we obtain a SFR for the extended emission of $ 1.32 \pm 0.37 \ M_\odot$ yr$^{-1}$, which is orders of magnitude lower than the SFR estimates for any of the SMGs in the field. This is consistent with the idea that most ($99\%$) of the UV radiation is obscured by dust within the SMGs.

%Similarly, we can use the models of \cite{cen13}, to obtain an estimate of the halo mass ($M_\text{h}$) of a central galaxy producing the extended Lyman-$\alpha$ emission, by: 
%\begin{equation}
%    L_{\text{Ly}\alpha}=10^{42.4} \Big( {\frac{M_\text{h}}{10^{12}M_\odot}}\Big)^{1.15}  \text{erg s}^{-1} 
%    \label{eq:2}
%\end{equation}

%For the LAB, we obtain a $M_h = 1.56 \pm 0.37 \times 10^{10} \ M_\odot$. This halo mass is three orders of magnitude smaller than the mass inferred for the protocluster core \citep{hill20}. A possible explanation for this discrepancy is that the LAB is relatively small compared with the typical values found in the literature \citep[$\sim100$ kpc][]{cantalupo17}, with an extension of only 70 kpc. If we compare with the results of \cite{cen13}, the majority of LABs reside in more evolved protoclusters at lower redshifts, with halo masses, of $\sim 10^{12} M_\odot$. Therefore, such scaling relation might not be applicable to the SPT2349-56 system. 

%This opens up a window to think also of the contribution of an AGN that could be hidden by dust and/or in an inactive phase. This last picture could be consistent with the idea that the DSFGs are a significant population in high redshift protoclusters \citep{chiang13,casey16}, where most of the star-forming activity in the DSFGs are happening almost simultaneously (possibly triggered by the denser environment in the protocluster).

Recent radio imaging of the SPT2349-56 field using the Australia Telescope Compact Array (ATCA) and the Australian Square Kilometer Array Pathfinder (ASKAP) found strong radio emission from the protocluster core complex (Chapman et al. in preparation). The steep radio spectrum found clearly indicates that at least one of the three central sources (B, C and G in the nomenclature used by Miller et al., or C3, C6 and C13 following Hill et al.) host a radio AGN. This finding supports the idea that enhanced Lyman-$\alpha$ emission at the LAB location is produced by AGN activity at the protocluster core \citep[e.g.][]{vito20}.

%Finally, we should also consider that this method to compute the halo mass might be severely biased due to the important dust obscuration that most of the galaxies in the SPT2349-56 system are subject. The vast amounts of dust hosted by these DSFGs will only permit a minor fraction of the ionizing photon to escape therefore limiting the interpretation of the Lyman-$\alpha$ emission flux. 

Figure \ref {fig:pv} shows a position-velocity (PV) diagram of the Lyman-$\alpha$ emission of the LAB, extracted along the x-axis (0 degrees of inclination) towards the west of the MUSE datacube, with a slit width of 14.2 arcsec. Here we note a widespread emission along the central velocity with an extension of 5 arcsec. At the edges, for the more distant structure, the emission goes towards bluer velocities. On the other hand, in the edge closer to the cluster, we have a structure that shifts to positive velocities. Another important issue is the behavior of the luminosity in the PV diagram. We can divide the structure into two different blobs: with the western being brighter than the eastern. This result is in agreement with the spectral line of the LAB, where we observe that the reddest emission is strongest and wider than the bluest emission (Figure \ref{fig:blobcii}).

\subsection{Modeling the Lyman-$\alpha$ spectrum of the extended emission}
\begin{table*}[ht]
\caption{Results from the radiative transfer modeling of the LAB line profile}              % title of Table
\label{tab:model}      % is used to refer this table in the text
\centering
\begin{tabular}{ccccccccc}
\hline\hline                        % inserts double horizontal lines
Model & Source$^\dagger$ & $z_{\rm sys}$ & $v_{\rm exp}$ & log($N_{\rm HI}$) & $\tau_{\rm d}$ & log($T$) & EW(Lyman-$\alpha$) & $\sigma$(Lyman-$\alpha$)$^\ddag$  \\    % table heading
 & & & (km s$^{-1}$) & (cm$^{-2}$) &  & (K) & ($\AA$)  & (km s$^{-1}$) \\
\hline

0 & C10 & $4.2895\pm0.0019$ & $480_{-14}^{+9}$ & $19.85_{-0.11}^{+0.17}$ & $0.01_{-0.01}^{+0.02}$ & $4.2_{-0.6}^{+0.4}$ & $5.1_{-1.2}^{+1.3}$ & $798_{-31}^{+15}$ \\ % LAB
1 & C14 & $4.3057\pm0.0020$ & $377_{-46}^{+28}$ & $16.19_{-0.22}^{+0.43}$ & $3.7_{-1.6}^{+0.9}$ & $3.1_{-0.2}^{+0.4}$ & $6.6_{-0.8}^{+0.9}$ & $328_{-22}^{+33}$ \\ % LAB
2 & C17 & $4.3049\pm0.0020$ & $375_{-31}^{+17}$ & $16.15_{-0.19}^{+0.43}$ & $2.8_{-1.3}^{+1.1}$ & $3.15_{-0.23}^{+0.35}$ & $6.6_{-0.8}^{+1.0}$ & $330_{-16}^{+30}$ \\ % LAB
3 & Core & $4.3040\pm0.0020$ & $310_{-22}^{+36}$ & $16.65_{-0.50}^{+0.39}$ & $1.7_{-1.2}^{+1.8}$ & $3.1_{-0.2}^{0.3}$ & $6.6_{-0.9}^{+1.1}$ & $346_{-35}^{+42}$ \\ % LAB
\hline
\end{tabular}
\flushleft {\bf Notes:} $^\dagger$ Source assumed to be producing the Lyman-$\alpha$ emission. Its [C\textsc{ii}] redshift is assumed to be the systemic redshift of the system for each model. $^\ddag$ $\sigma=$ FWHM/2.35. 
\end{table*} 
\begin{figure}
    \centering
    \includegraphics[scale=0.5]{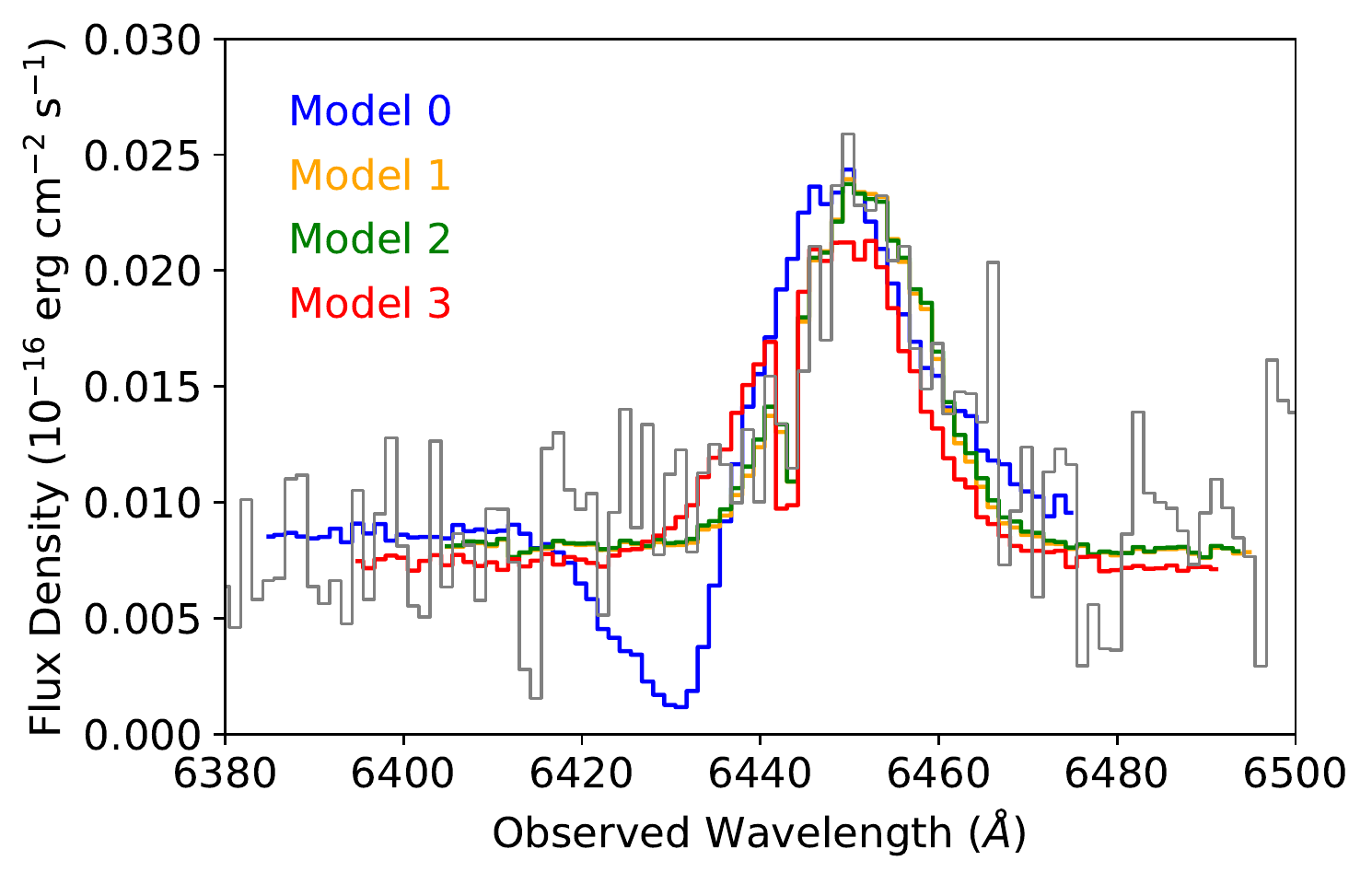}
    \caption{Lyman-$\alpha$ spectrum of the LAB compared to the best-fit models that assume different systemic redshifts for the emitting source. The models are described in the text and their best-fit parameters are listed in Table \ref{tab:model}. The best description of the observed Lyman-$\alpha$ spectrum is given by models 1-3, suggesting that the emission is produced by photoionization from either DSFGs C14, C17 or the protocluster core.}
    \label{fig:model}
\end{figure}

To further explore the origin of the LAB, we test the idea that either the protocluster core or the galaxies spatially coincident with the LAB are the source of the Lyman-$\alpha$ emission. 

For this, we use the Lyman-$\alpha$ line profile of the LAB and the  Lyman-$\alpha$ Monte Carlo Radiate Transfer code \texttt{tlac} \citep{gronke14a, gronke15}. We utilize a expanding shell model, which has been widely used in several studies to successfully reproduce the Lyman-$\alpha$ profiles of galaxies in different redshifts and environments. The model assumes an homogeneous, spherical shell that expands radially outwards, with uniformly mixed  neutral gas (HI) and dust \citep{verhamme06}, and the emitting source located at the center of the shell. The shell model is defined by a set of seven parameters including the expanding velocity ($v_{\rm exp}$), the HI column density ($N_{\rm HI}$), the dust optical depth ($\tau_{\rm d}$), the effective temperature of the gas ($T$), the systemic redshift of the emitter ($z_{\rm sys}$), the intrinsic equivalent width of the Lyman-$\alpha$ line (EW(Lyman-$\alpha$)) and the intrinsic FWHM of the Lyman-$\alpha$ line (FWHM(Lyman-$\alpha$)). For more details on these parameters, we refer the reader to \citet{gronke15}. Given a constraint for the systemic redshift of the source and the input Lyman-$\alpha$ spectrum, the code yields the most likely set of parameters that reproduce the observed spectrum under the assumed geometry.

Since we are interested in learning which of the underlying starburst galaxies might be producing the Lyman-$\alpha$ emission, we constrain $z_{\rm sys}$ using the [C\textsc{ii}]-based redshifts of each of the possible sources of the Lyman-$\alpha$ line: C10 (model $0$), C14 (model $1$) and C17 galaxies (model $2$), and the protocluster core (model $3$). Instead of simply fixing $z_{\rm sys}$, we allow for a range in redshift given by the $3\sigma$ uncertainty around measured [C\textsc{ii}] redshift in each case. Since these ranges overlap, some of the solutions found are similar between each other. The results of this procedure are shown in Figure \ref{fig:model} and listed in Table \ref{tab:model}. 

We find that the model $0$ does not converge into a proper fit to the data, mostly due to the significant difference between the [C\textsc{ii}] redshift and that of the Lyman-$\alpha$ line. This forces the model to a high expansion velocity ($\sim500$ km s$^{-1}$) and low dust optical depth. This solution is less preferred, since the assumed emitting source (the C10 galaxy) is a gas-rich, dusty galaxy contrary to the result of a low dust optical depth.

The best fits are produced when using higher systemic redshifts (models $1-3$), which are more consistent with the redshift of the Lyman-$\alpha$ line. This is the case for sources C14 and C17, and the protocluster core. In these cases, the solutions are similar, yielding high outflow velocities ($\sim300-400$ km s$^{-1}$) and yet very low HI column densities (log(N(HI))$\sim16$). In these cases, the opacities appear to be moderate ($\tau>1.5-3.0$), yet more consistent with the dusty nature of the purported emission sources. Based on these results alone it is hard to disentangle the origin of the LAB. However, if the protocluster core starbursting galaxies are producing the Lyman-$\alpha$ emission it would require a complex patchy geometry where some of the UV radiation escapes and illuminates the HI gas in the LAB direction. While this is a plausible scenario, supported by the moderate optical depth of this solution (model 3), such solution is less likely than the scenario where the UV radiation is produced in-situ by either the C14, the C17 an/or both galaxies.

\section{Summary and conclusions} \label{sec:summ}

We presented a census of Lyman-$\alpha$ emission toward the IR-bright protocluster SPT2349-56 at $z=4.3$ obtained using MUSE observations. Through a blind search of Lyman-$\alpha$ emission towards the protocluster core and northern extension, we found three LAEs at distances $ > 90$ kpc from the protocluster core. The LAEs are bound to the $9 \times  10^{12} M_\odot$ protocluster core and all of them are redshifted relative to SPT2349-56. Only one of the ALMA SMGs previously identified in this field is tentatively detected in Lyman-$\alpha$.

Using a continuum-subtracted narrowband image we detect extended Lyman-$\alpha$ emission, which we refer to as a LAB, with a size of about 70 kpc across, located at $\sim 56$ kpc to the east of the protocluster core. The bulk of the LAB emission is also redshifted with respect to the core of the protocluster, in agreement with a red-skewed asymmetric profile. 

Two of the spatially overlapping DSFGs C14 and C17, are found to also coincide  spectrally, when comparing their [C\textsc{ii}] emission lines with that of the Lyman-$\alpha$ emission from the LAB \citep[][]{miller18,hill20}. This observation could be explained by the high star-formation activity seen in the DSFG protocluster members. Based on their locations and redshifts, the main suspects to be producing the ionizing photons and thus the Lyman-$\alpha$ emission are the C14 and C17 DSFGs, or the protocluster core. In the later case, the geometry of the dust distribution should allow the Lyman-$\alpha$ photons to get scattered from the core such that the photons find a region to escape to the east. Such scenarios are supported by radiative transfer modeling of the Lyman-$\alpha$ line profile of the LAB.

We do not find an overdensity of LAEs, or a source density comparable to what we might have expected from the number of [CII] and submillimeter continuum sources found in this field. We interpret this as a structure that is still heavily dust obscured and dominated by submm-detected galaxies. 

%\textcolor{red}{On the other hand, this could be explained by an AGN in an off-phase, where this could be responsible for the starbursting activity inside the protocluster members.  Summarizing this we could think that at high redshift DSFGs are a significant tracer of galaxies overdensities.}

\begin{acknowledgements}

This paper makes use of the
following ALMA data: ADS/JAO.ALMA\#2017.1.00273.S; and
ADS/JAO.ALMA\#2018.1.00058.S. ALMA is a partnership of ESO
(representing its member states), NSF (USA) and NINS (Japan), together with NRC (Canada), MOST and ASIAA (Taiwan), and KASI
(Republic of Korea), in cooperation with the Republic of Chile. The Joint ALMA Observatory is operated by ESO, AUI/NRAO, and
NAOJ.

Y.A. acknowledges partial support from Comit\'e Mixto ESO - Gobierno de Chile. MA acknowledges support from FONDECYT grant 1211951, CONICYT + PCI + INSTITUTO MAX PLANCK DE ASTRONOMIA MPG190030 and CONICYT+PCI+REDES 190194. This work was partially funded by the ANID BASAL project FB210003.  T.A. acknowledges support from the Millennium Science Initiative ICN12\_009. D.N. acknowledges support from the US NSF under grant 1715206 and Space Telescope Science Institute under grant AR-15043.0001. 
      
J.D.V. and S.J. acknowledge support from the US NSF under grants AST-1715213 and AST-1716127. S.J. acknowledge support from the US NSF NRAO under grants SOSPA5-001 and SOSPA7-006, and SOSPA4-007, respectively. J.D.V. acknowledges support from an A. P. Sloan Foundation Fellowship. E.J.J. acknowledges support from FONDECYT Iniciaci\'on en investigaci\'on 2020 Project 11200263.

\end{acknowledgements}

%-------------------------------------------------------------------
\bibliographystyle{aa}
\bibliography{Ly-alpha_tomography.bib}
%------------------------
\onecolumn

\begin{appendix}

\section{Lyman-$\alpha$ spectra toward the SPT2349-56 DSFGs}

The following figures show the observed MUSE spectra toward all the DSFGs in the SPT2349-56 system, centered at the expected location of the Lyman-$\alpha$ line emission.

\begin{figure*}[ht!]
    \centering
    \begin{subfigure}
        \centering
        \includegraphics[width=0.3\linewidth]{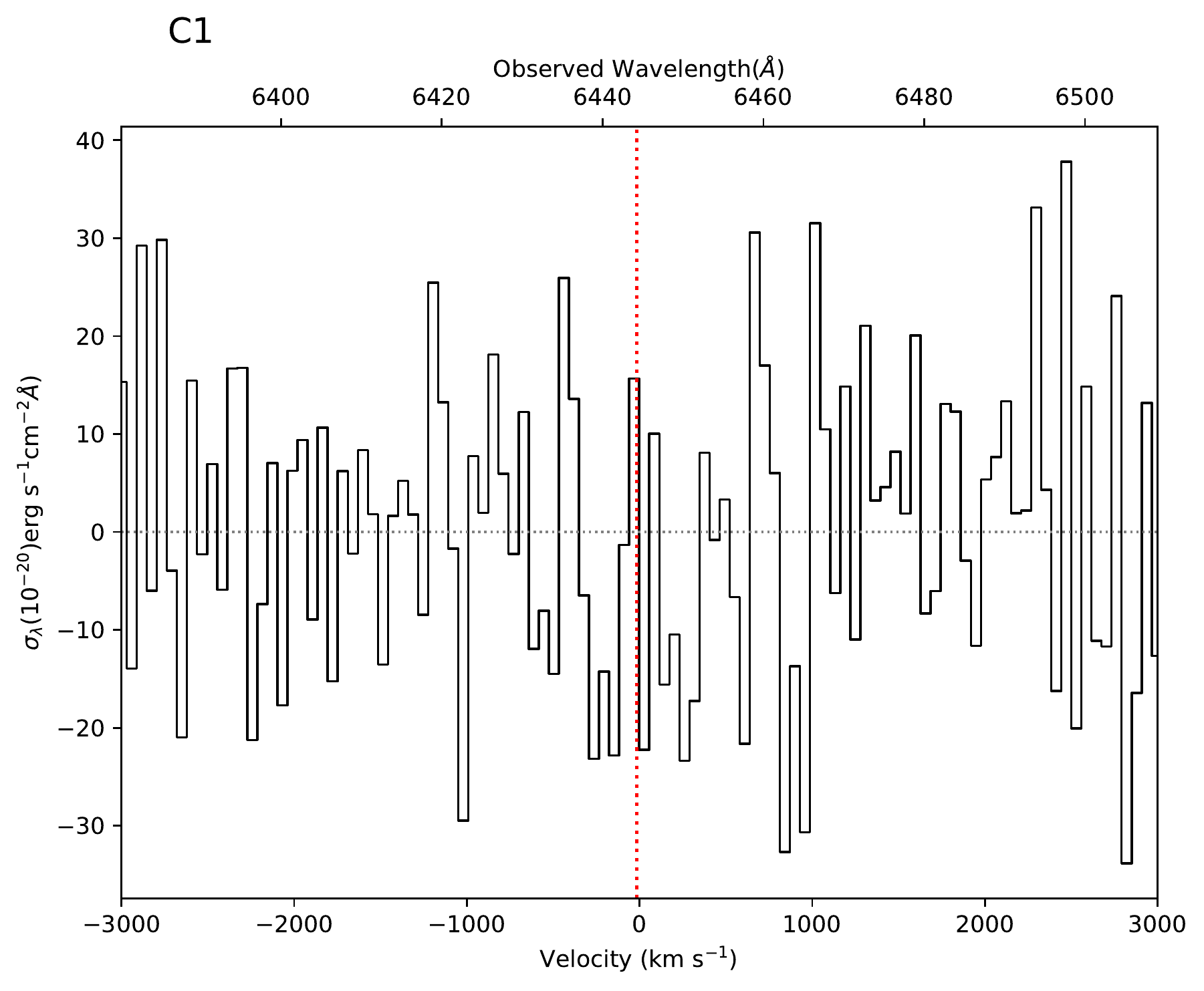} 
    \end{subfigure}
    \hfill
    \begin{subfigure}
        \centering
        \includegraphics[width=0.3\linewidth]{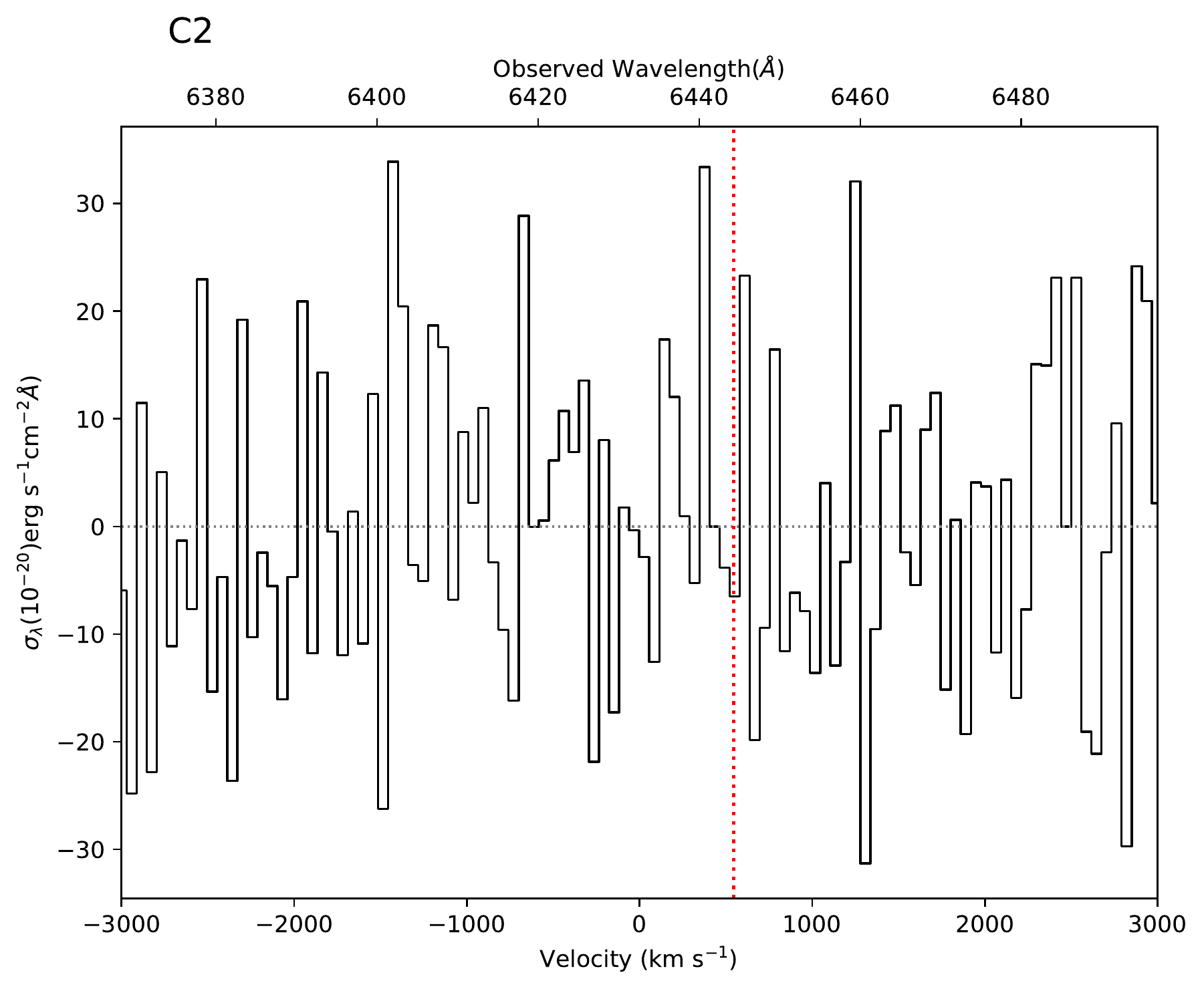} 
    \end{subfigure}
    \hfill
    \begin{subfigure}
        \centering
        \includegraphics[width=0.3\linewidth]{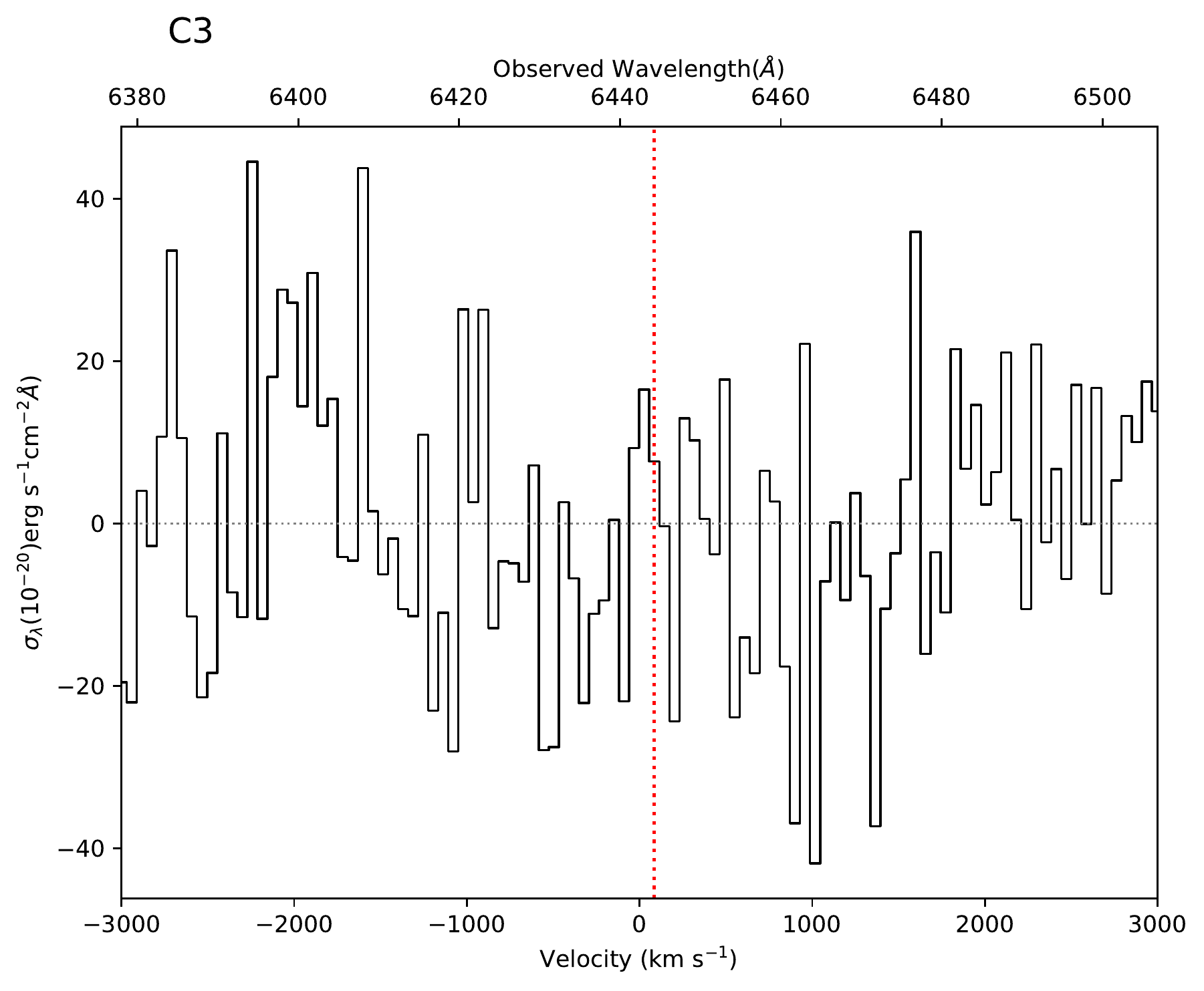} 
    \end{subfigure}
    \begin{subfigure}
        \centering
        \includegraphics[width=0.3\linewidth]{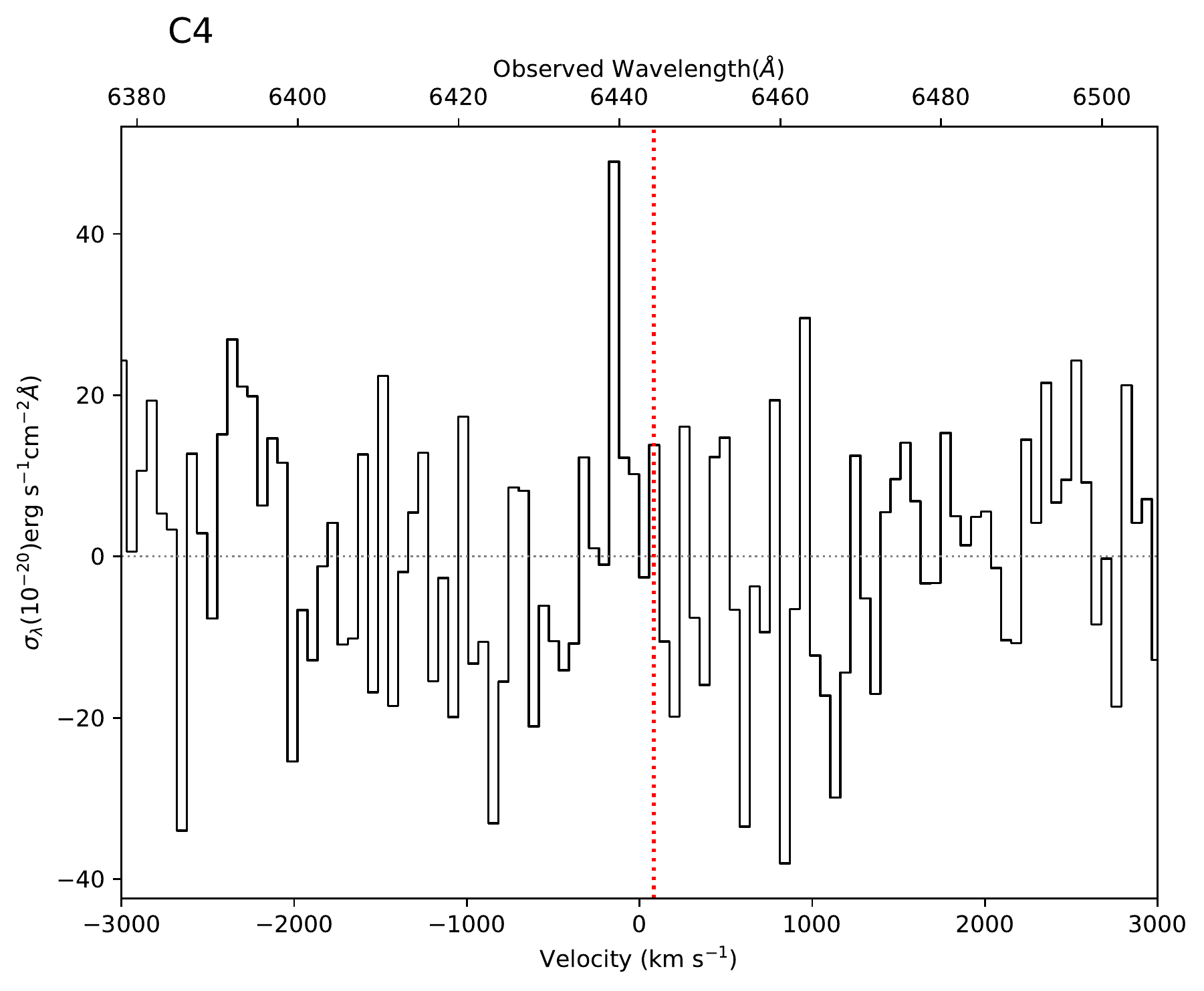} 
    \end{subfigure}
    \hfill
    \begin{subfigure}
        \centering
        \includegraphics[width=0.3\linewidth]{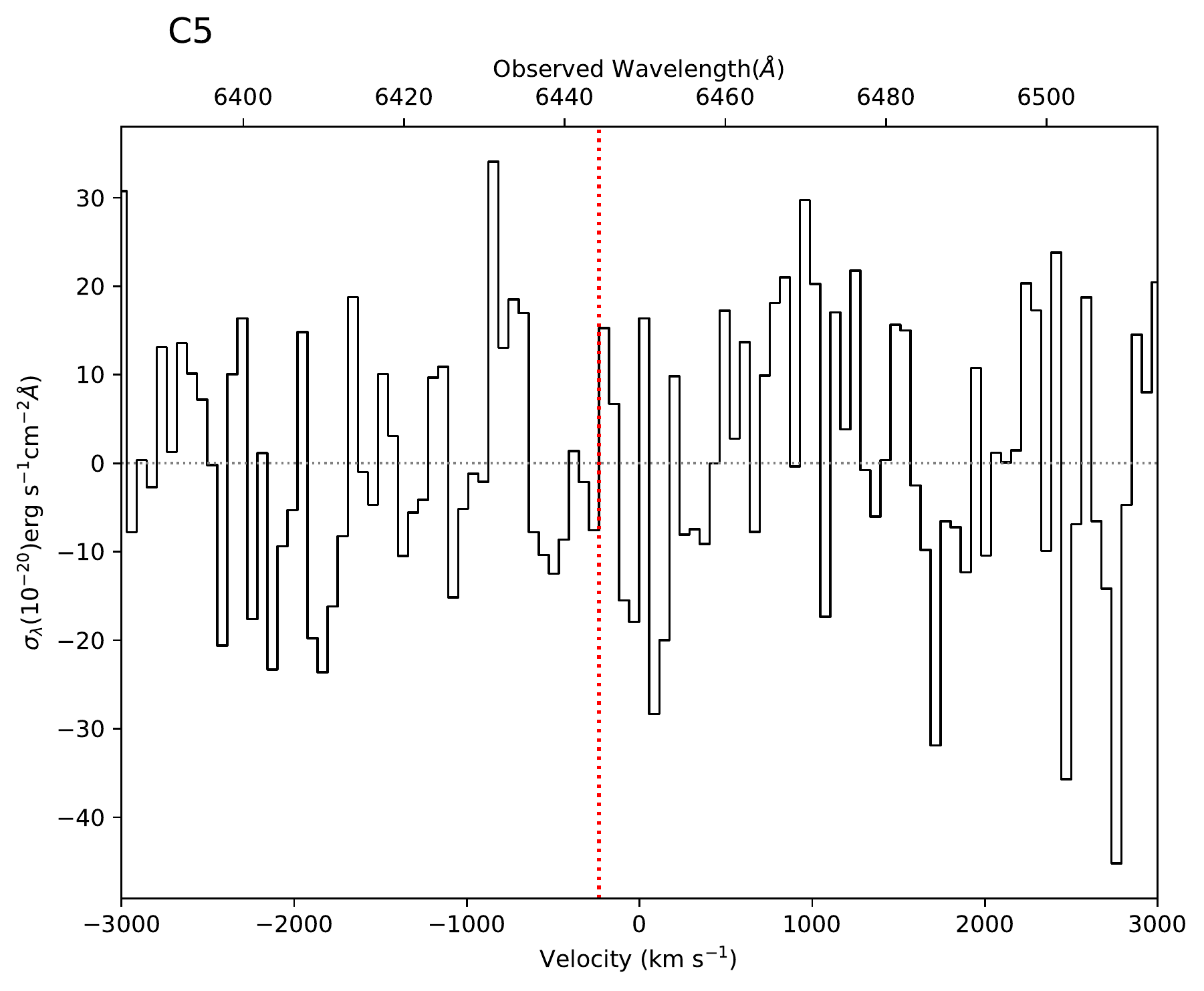} 
    \end{subfigure}
    \hfill
    \begin{subfigure}
        \centering
        \includegraphics[width=0.3\linewidth]{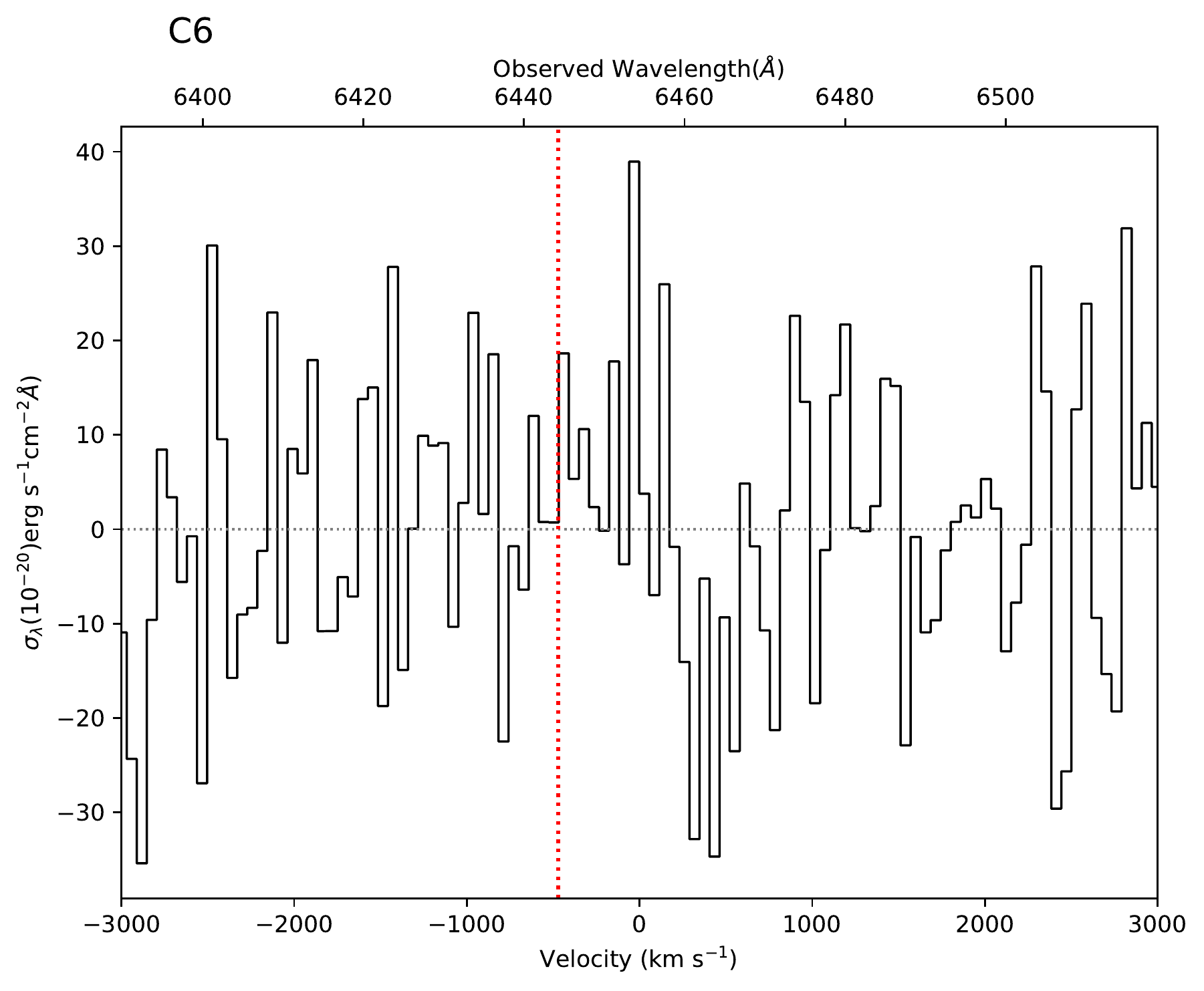} 
    \end{subfigure}
    \begin{subfigure}
        \centering
        \includegraphics[width=0.3\linewidth]{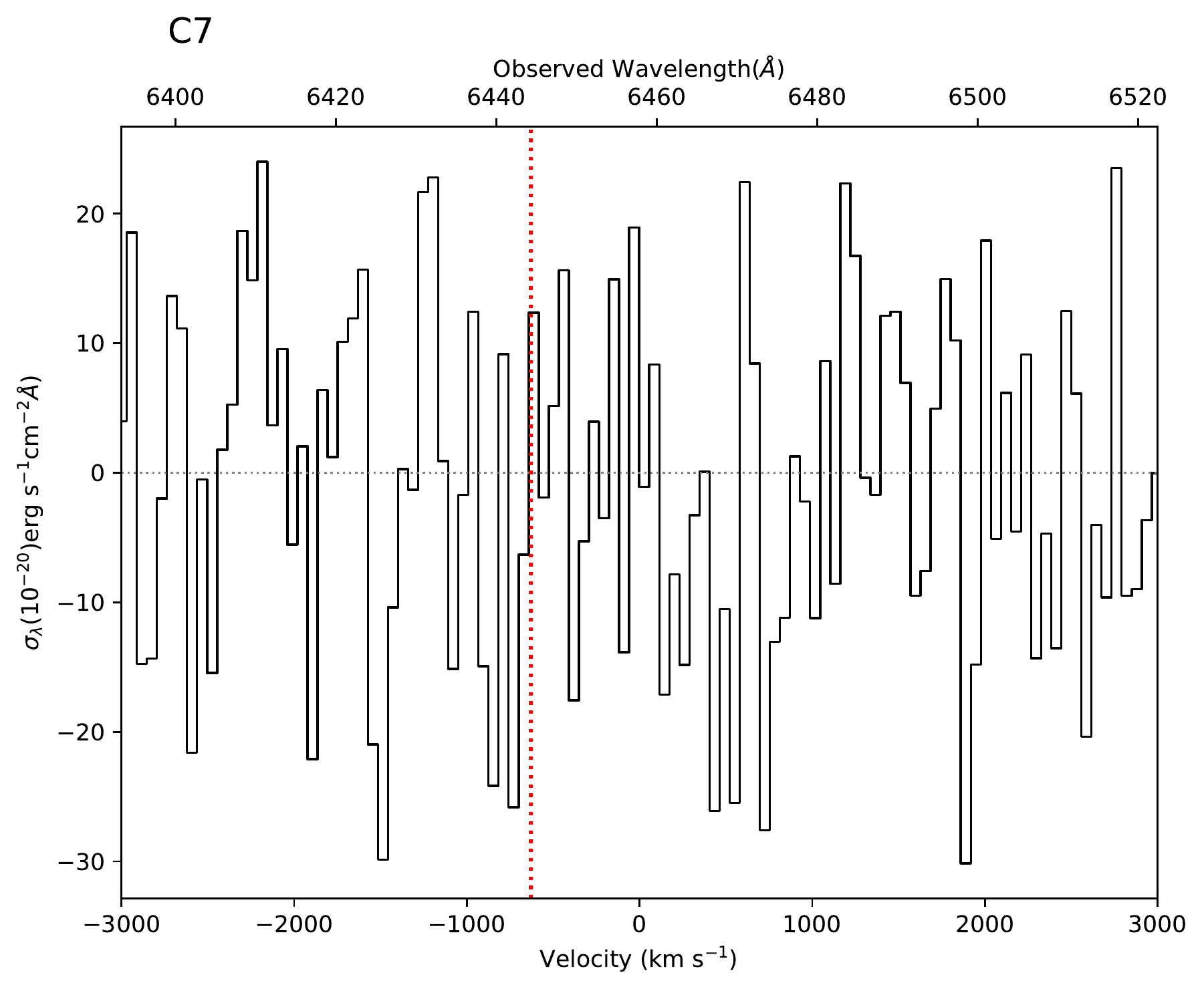} 
    \end{subfigure}
    \hfill
    \begin{subfigure}
        \centering
        \includegraphics[width=0.3\linewidth]{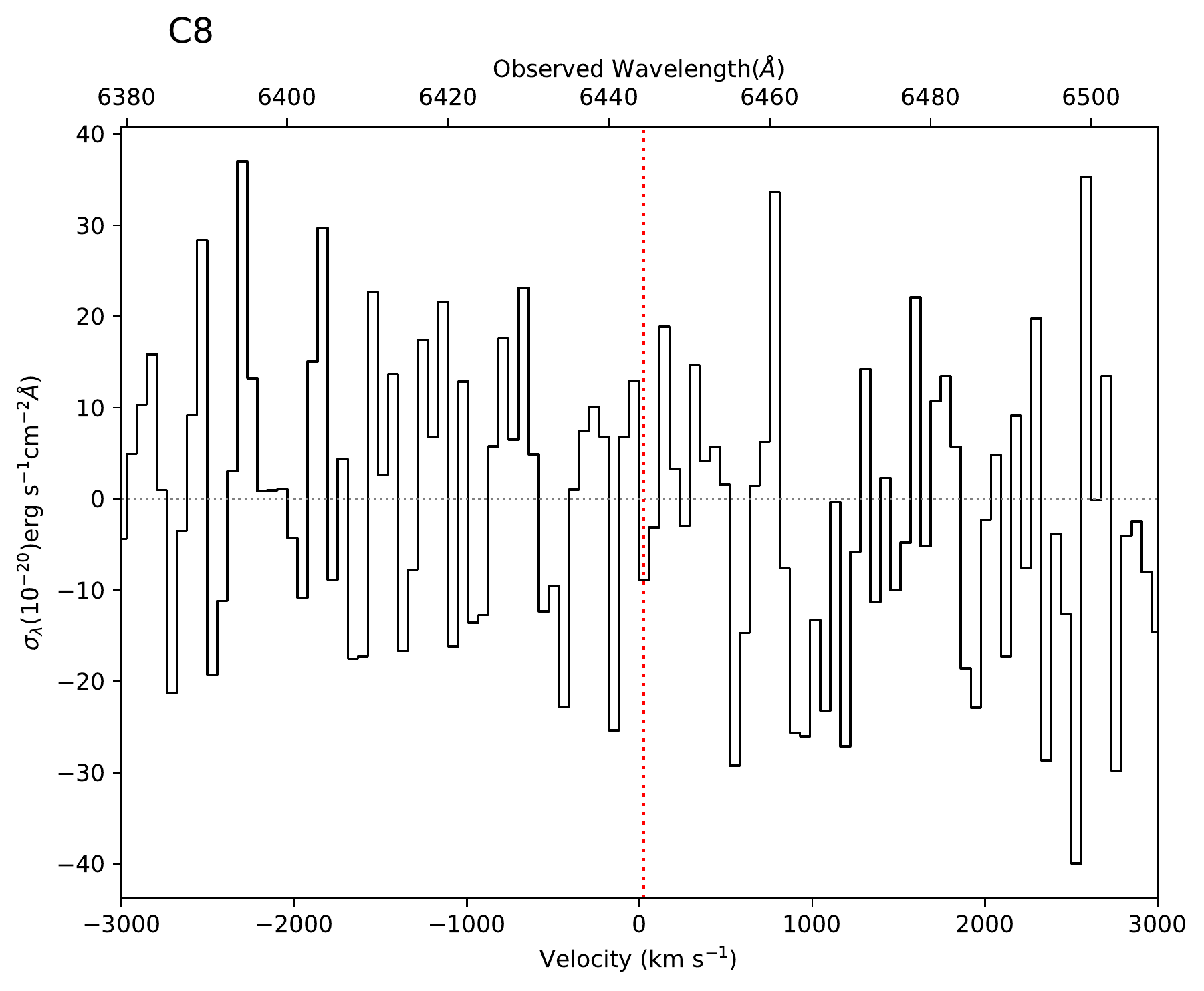} 
    \end{subfigure}
    \hfill
    \begin{subfigure}
        \centering
        \includegraphics[width=0.3\linewidth]{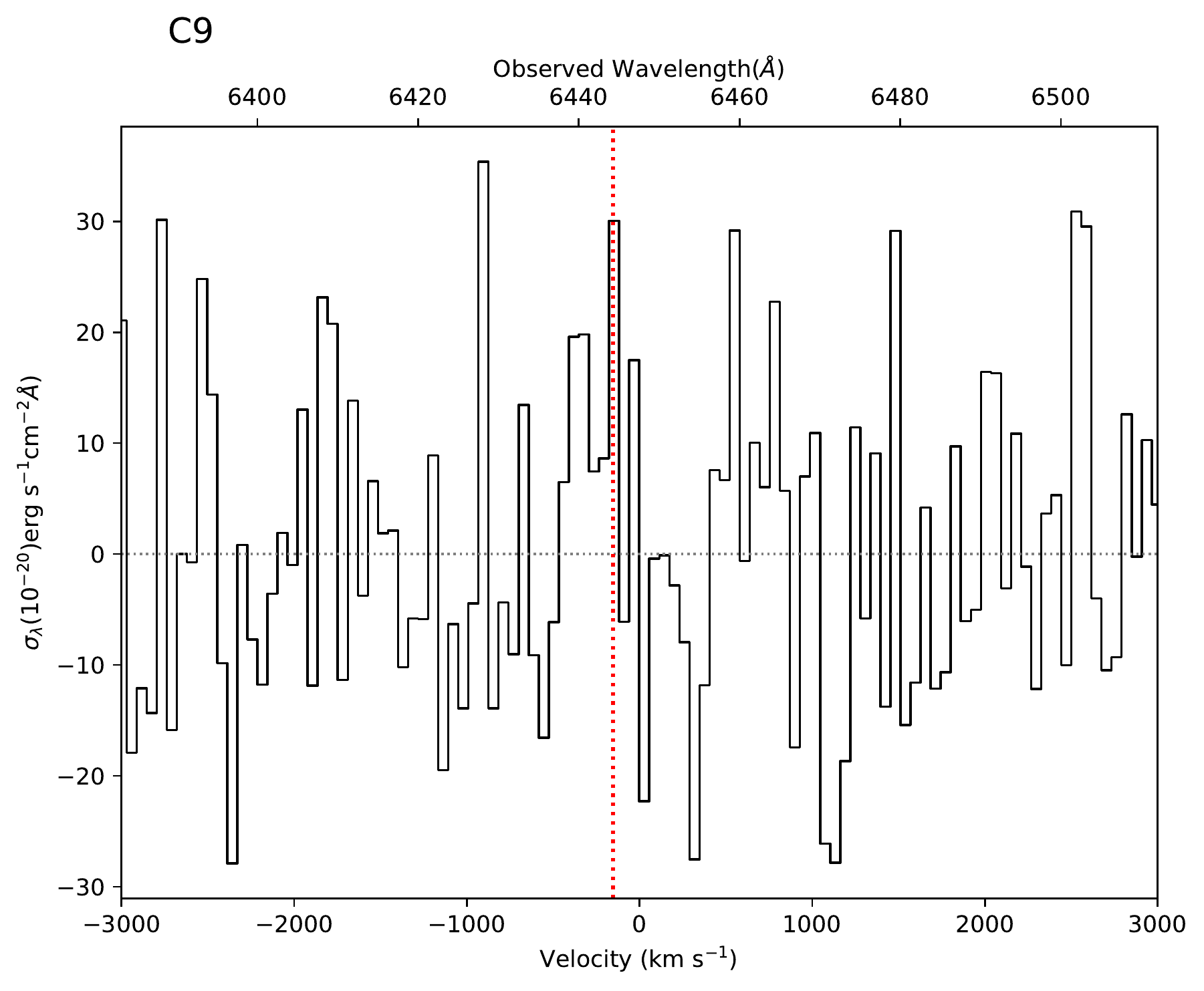} 
    \end{subfigure}
    \vspace{1cm}
    \begin{subfigure}
        \centering
        \includegraphics[width=0.3\linewidth]{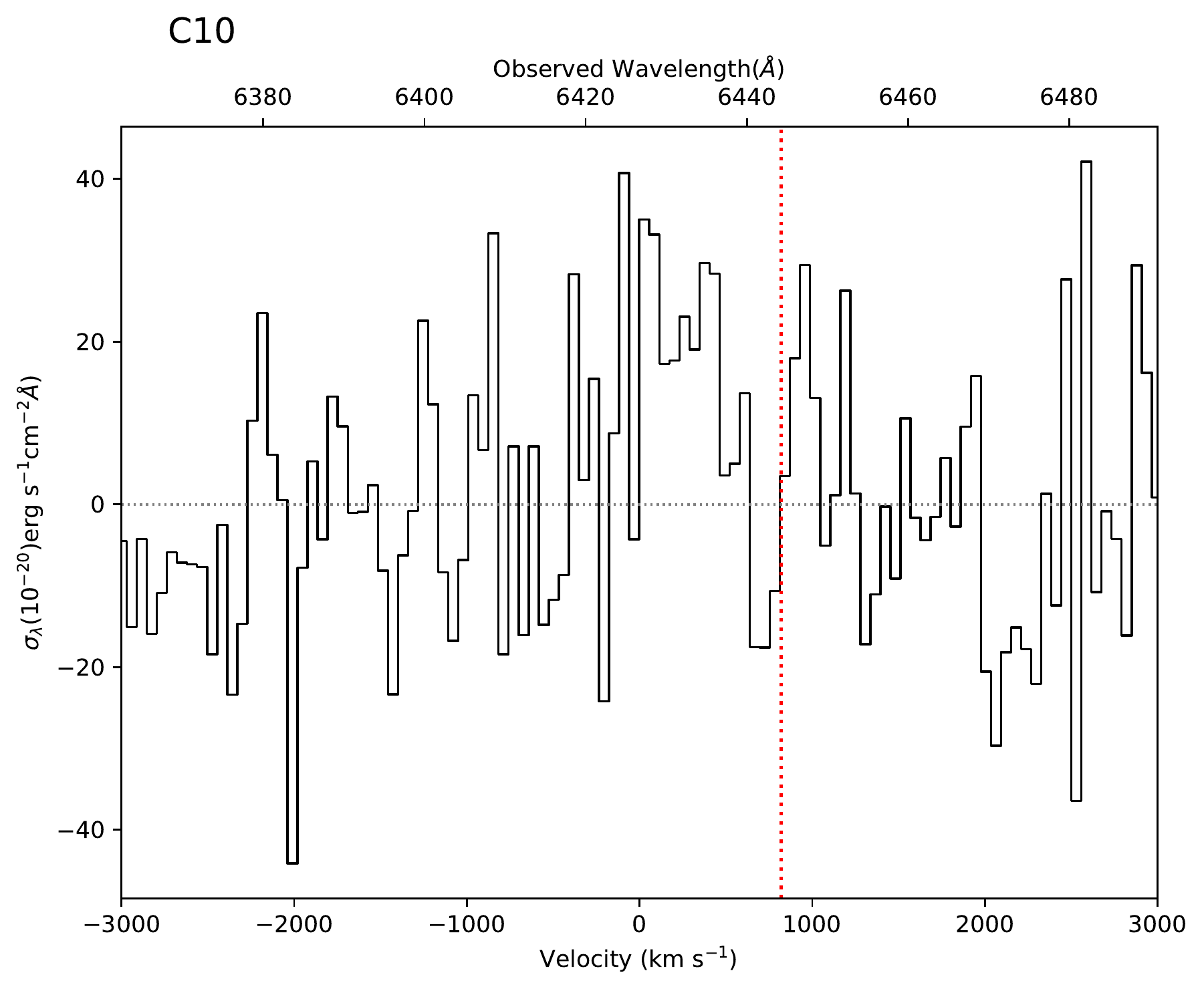} 
    \end{subfigure}
    \hfill
    \begin{subfigure}
        \centering
        \includegraphics[width=0.3\linewidth]{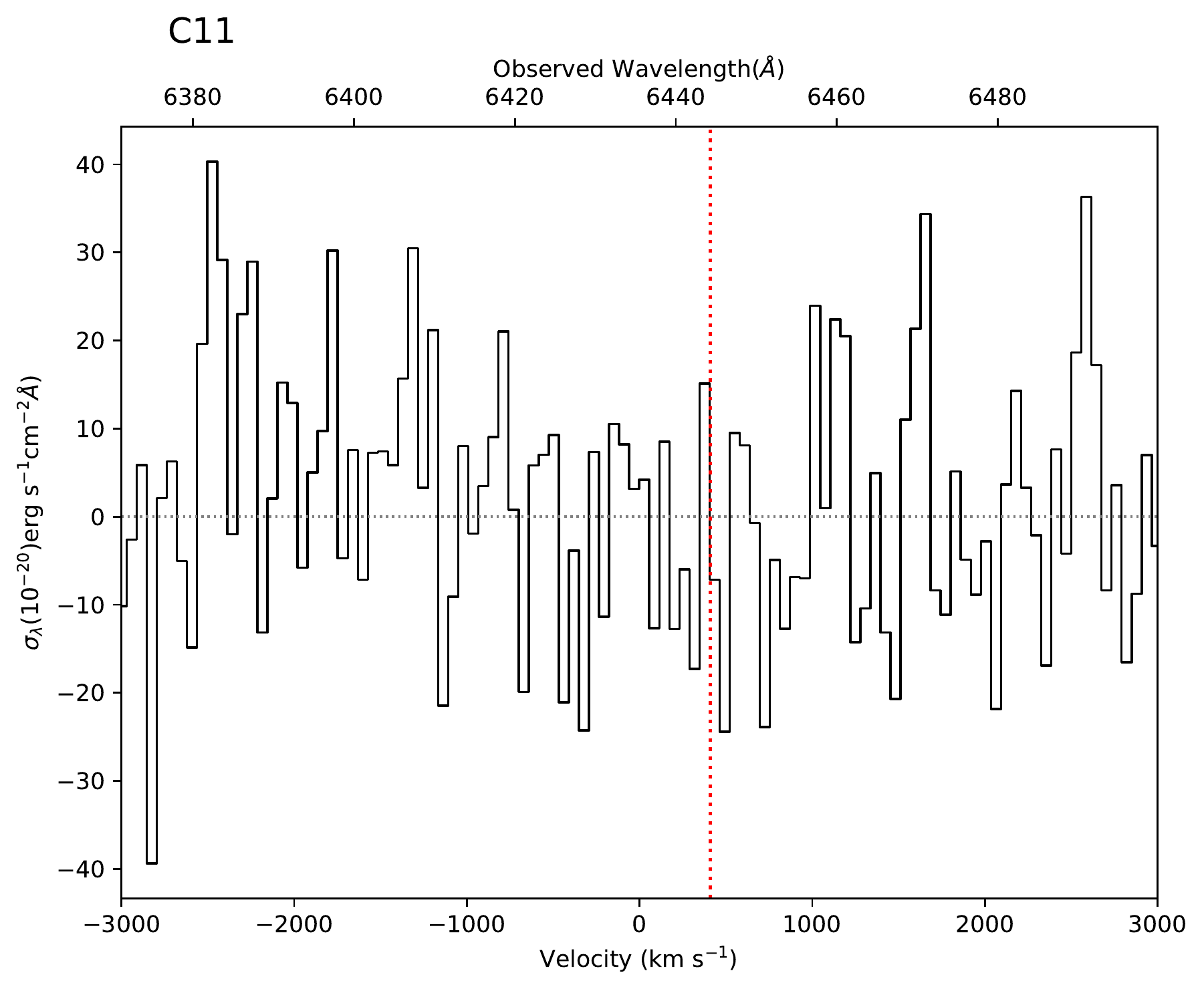} 
    \end{subfigure}
    \hfill
    \begin{subfigure}
        \centering
        \includegraphics[width=0.3\linewidth]{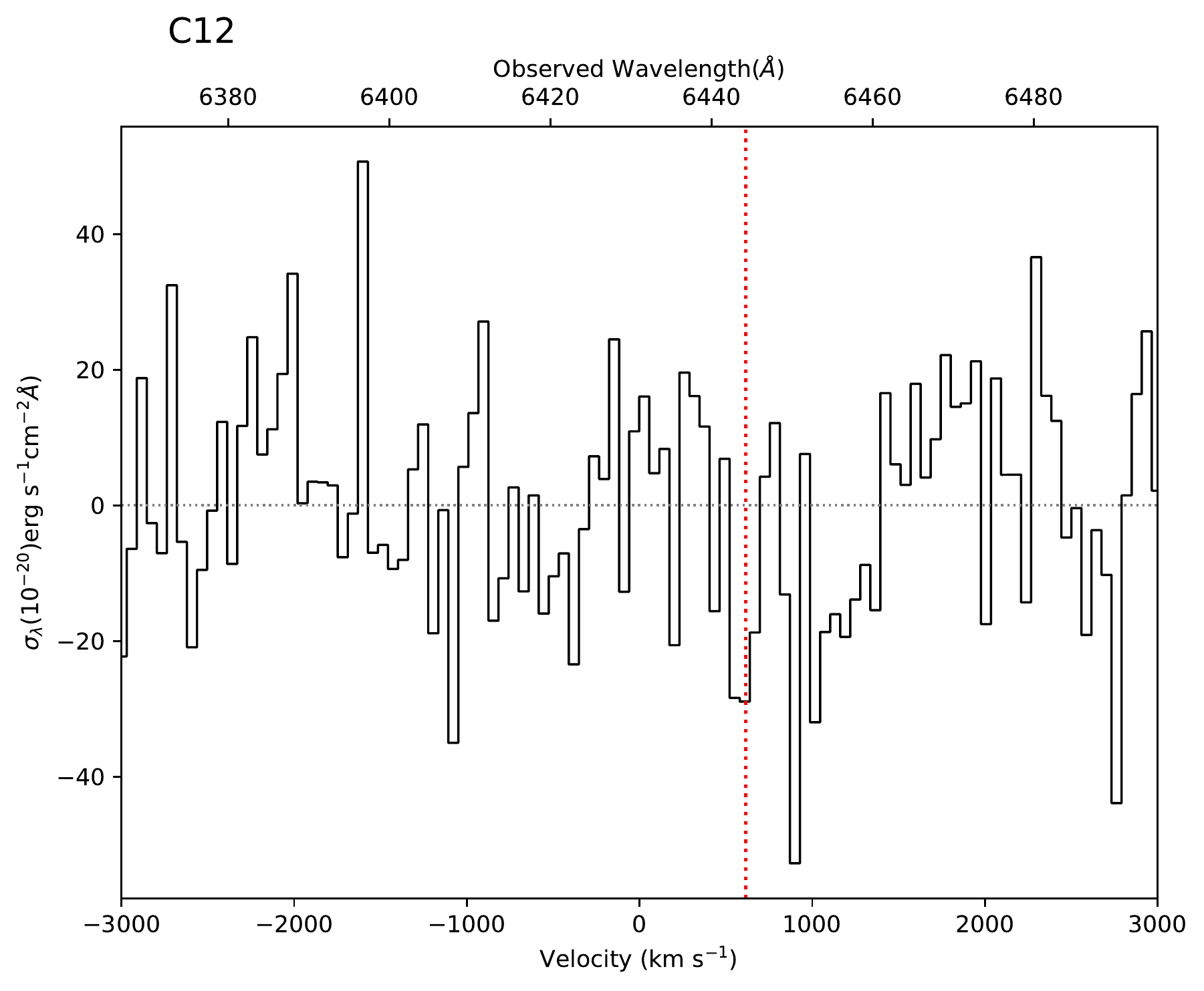} 
    \end{subfigure}
    
\end{figure*}

\begin{figure*}[ht!]
\begin{subfigure}
        \centering
        \includegraphics[width=0.3\linewidth]{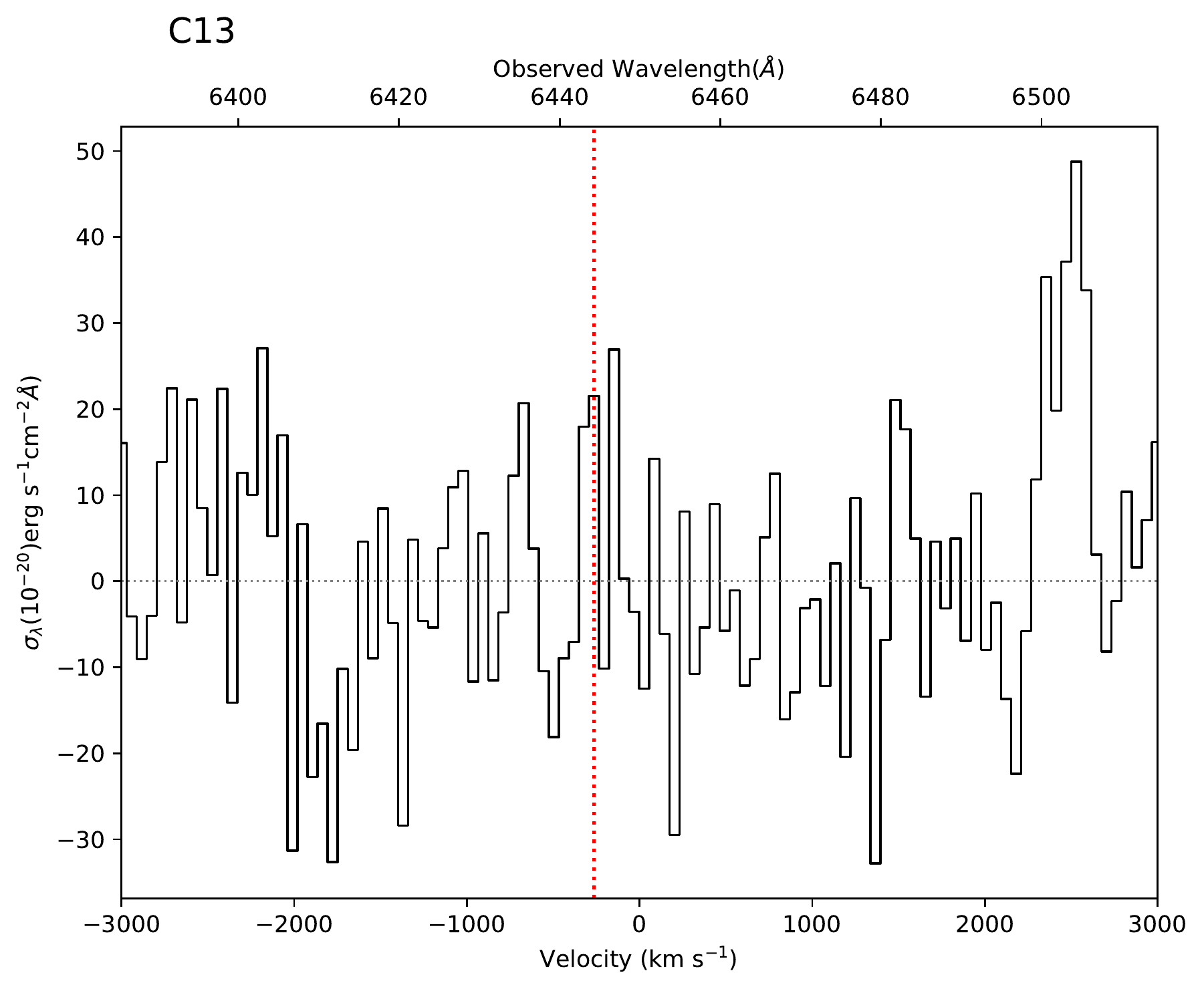} 
    \end{subfigure}
    \hfill
    \begin{subfigure}
        \centering
        \includegraphics[width=0.3\linewidth]{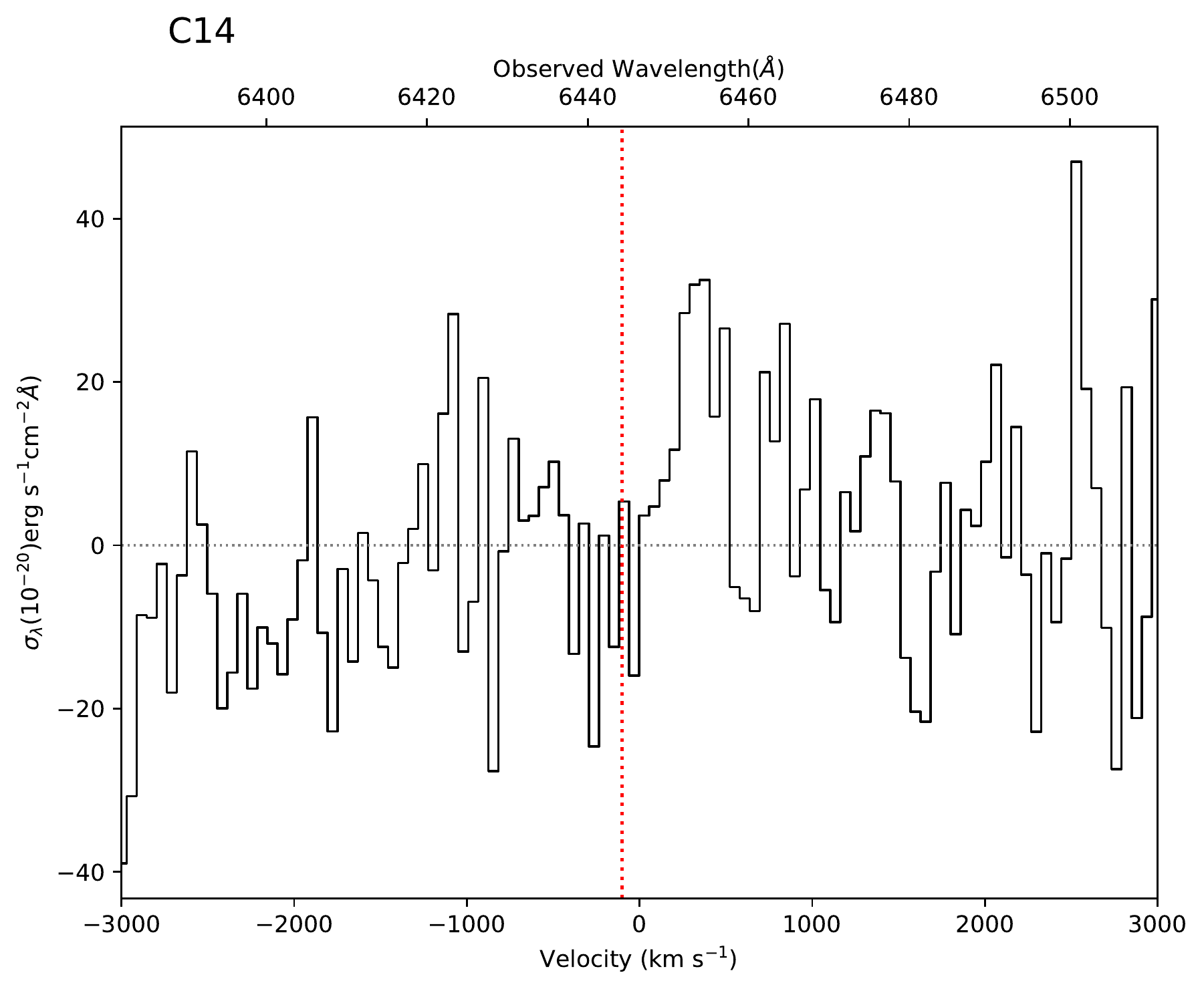} 
    \end{subfigure}
    \hfill
    \begin{subfigure}
        \centering
        \includegraphics[width=0.3\linewidth]{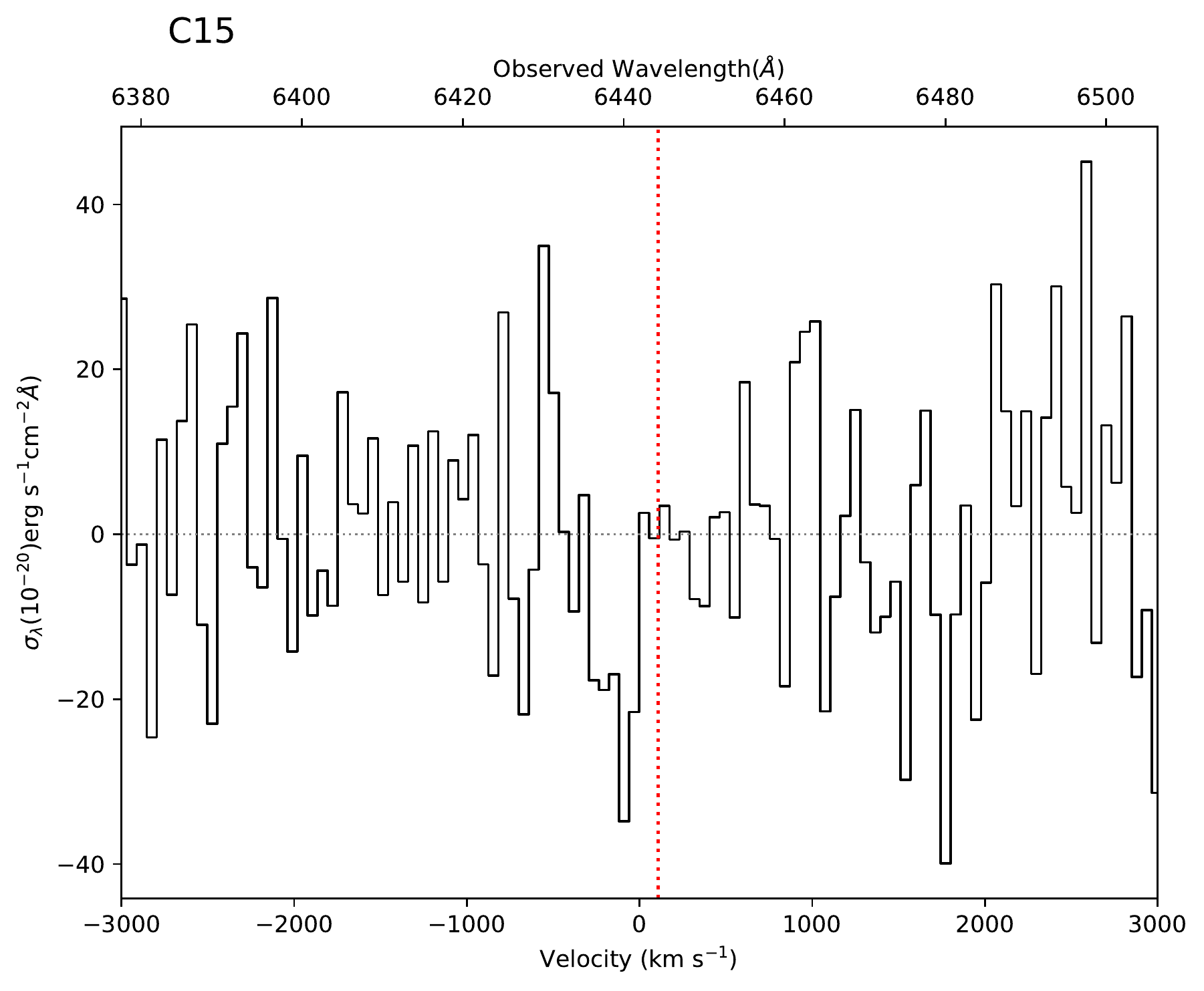} 
    \end{subfigure}
    \hfill
    \begin{subfigure}
        \centering
        \includegraphics[width=0.3\linewidth]{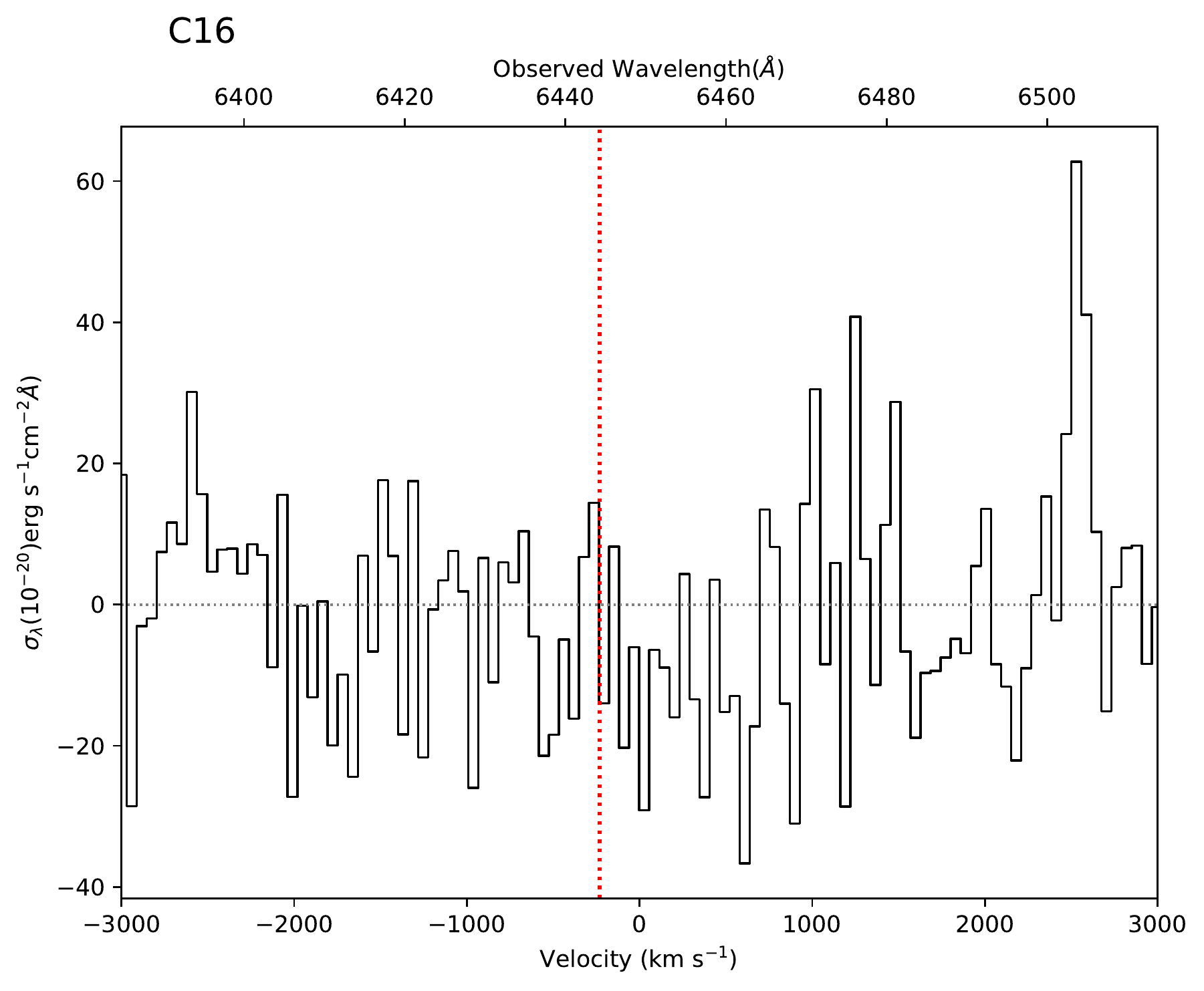} 
    \end{subfigure}
    \hfill
    \begin{subfigure}
        \centering
        \includegraphics[width=0.3\linewidth]{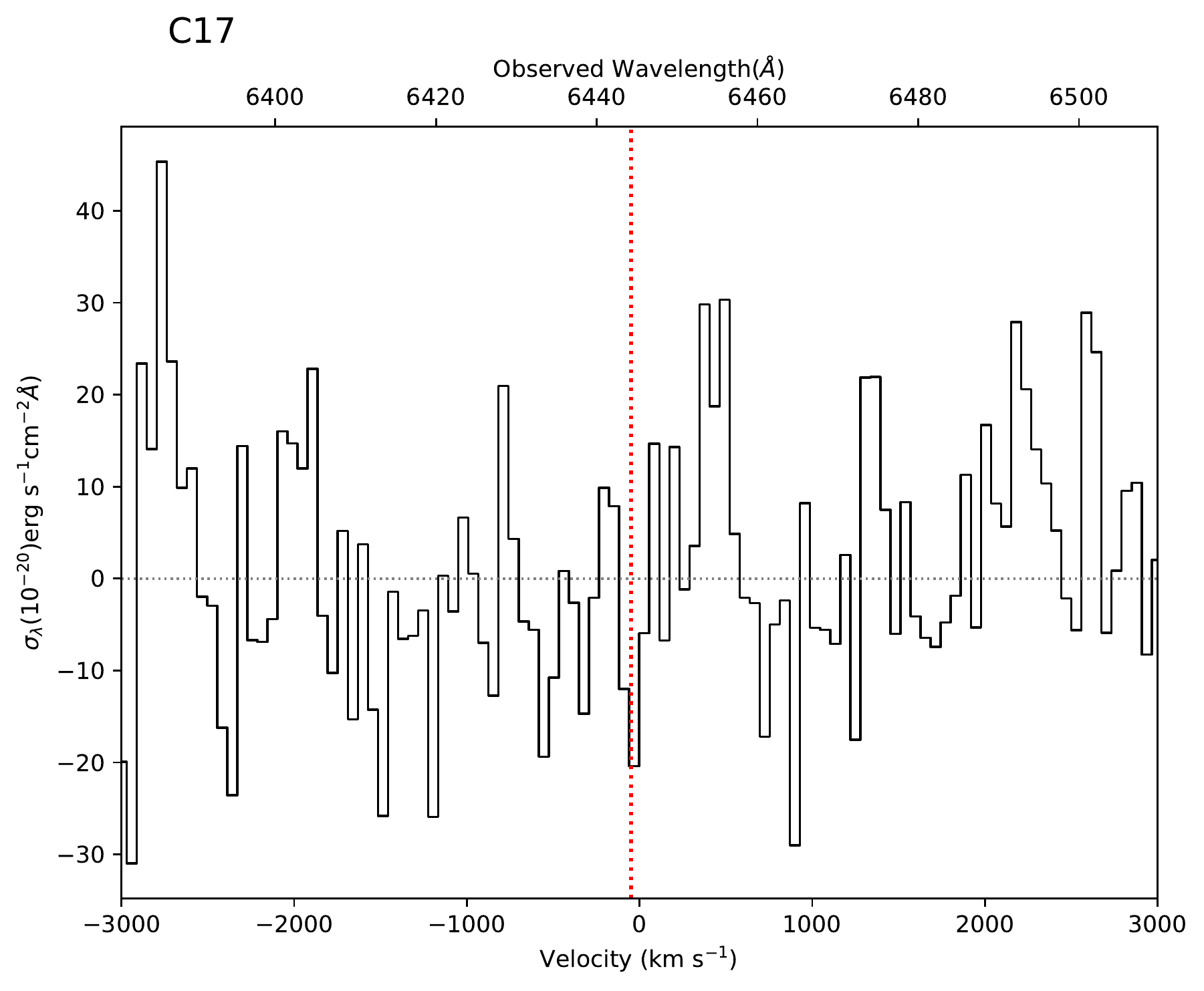} 
    \end{subfigure}
    \hfill
    \begin{subfigure}
        \centering
        \includegraphics[width=0.3\linewidth]{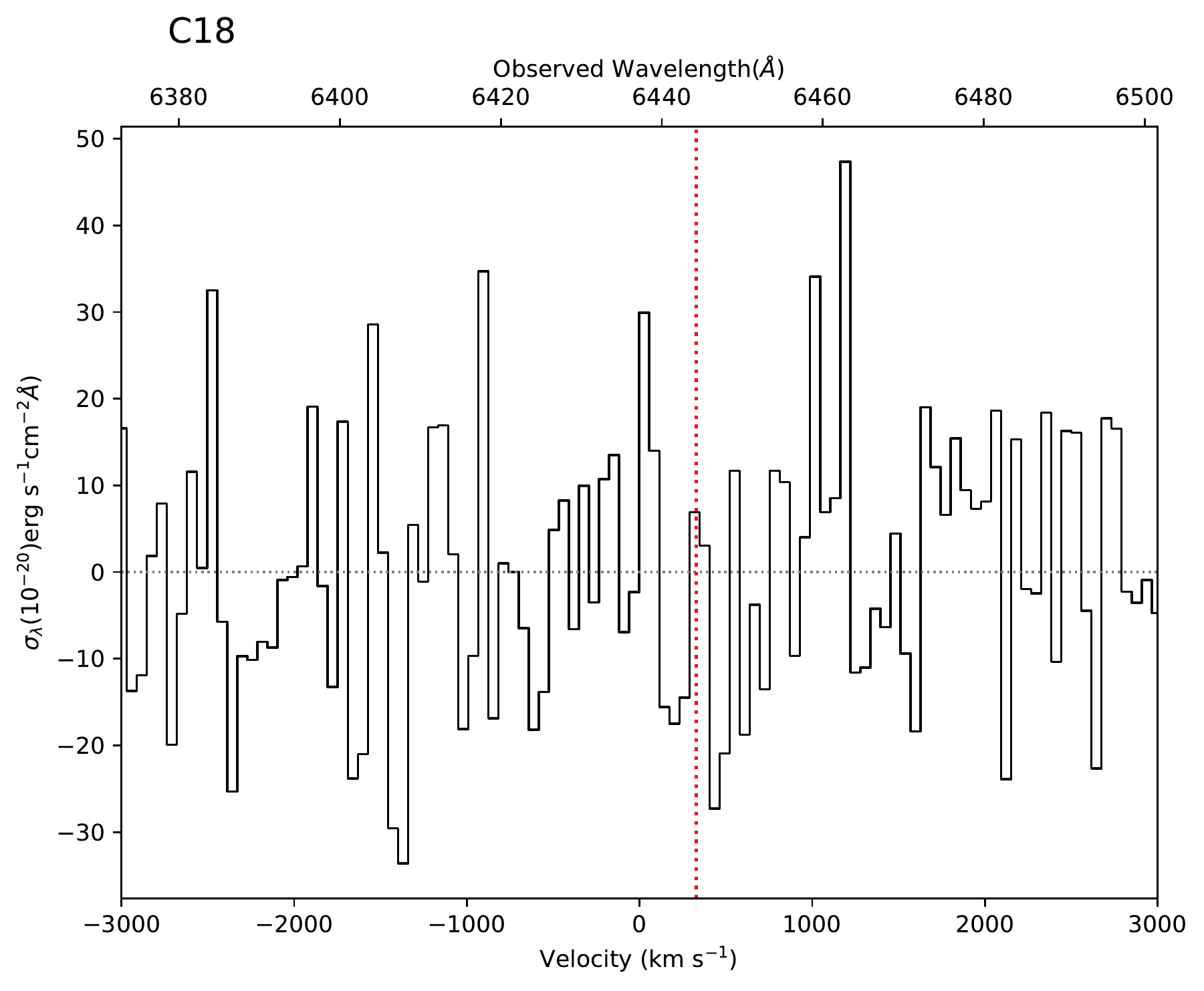} 
    \end{subfigure}
    \begin{subfigure}
        \centering
        \includegraphics[width=0.3\linewidth]{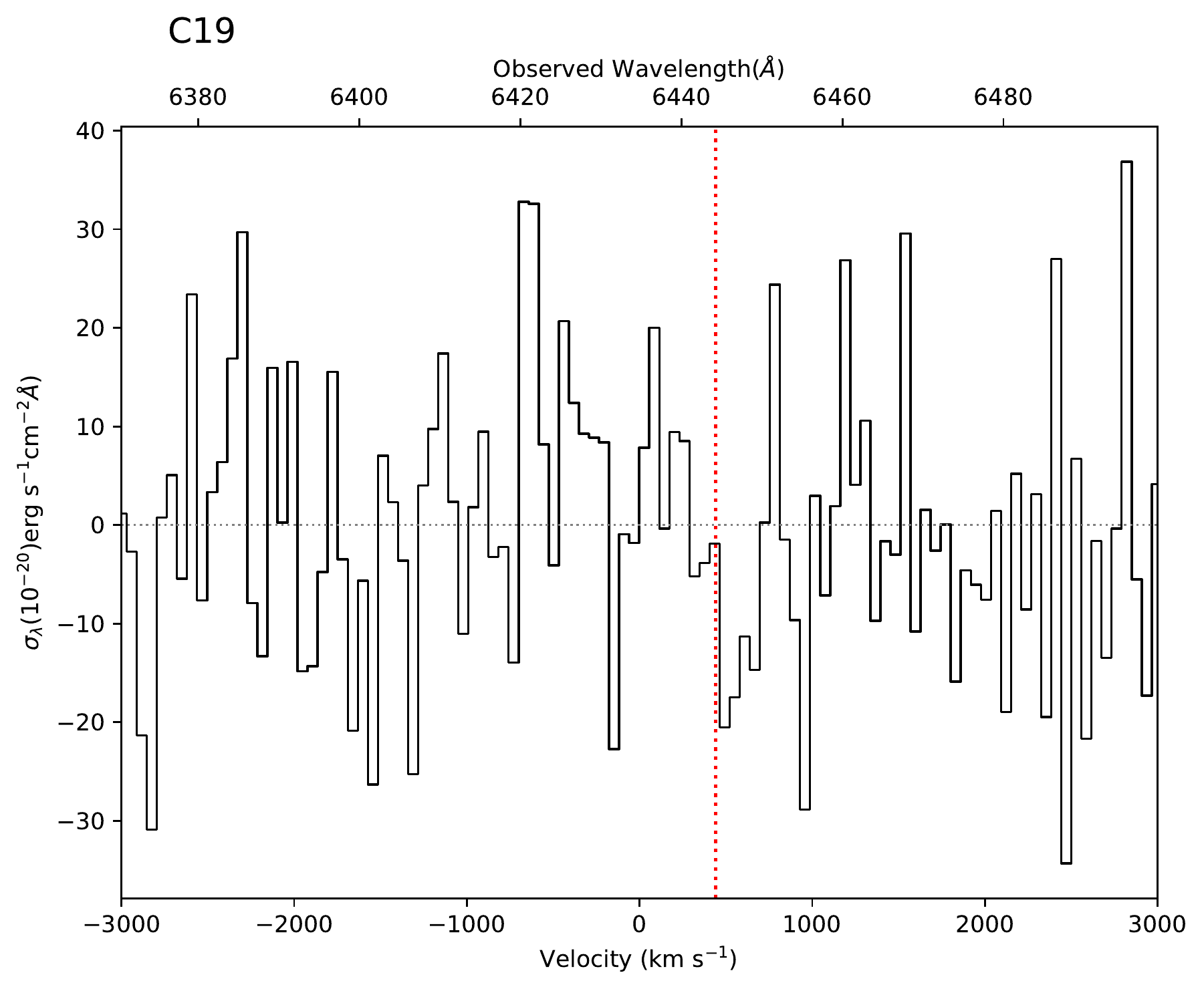} 
    \end{subfigure}
    \hfill
    \begin{subfigure}
        \centering
        \includegraphics[width=0.3\linewidth]{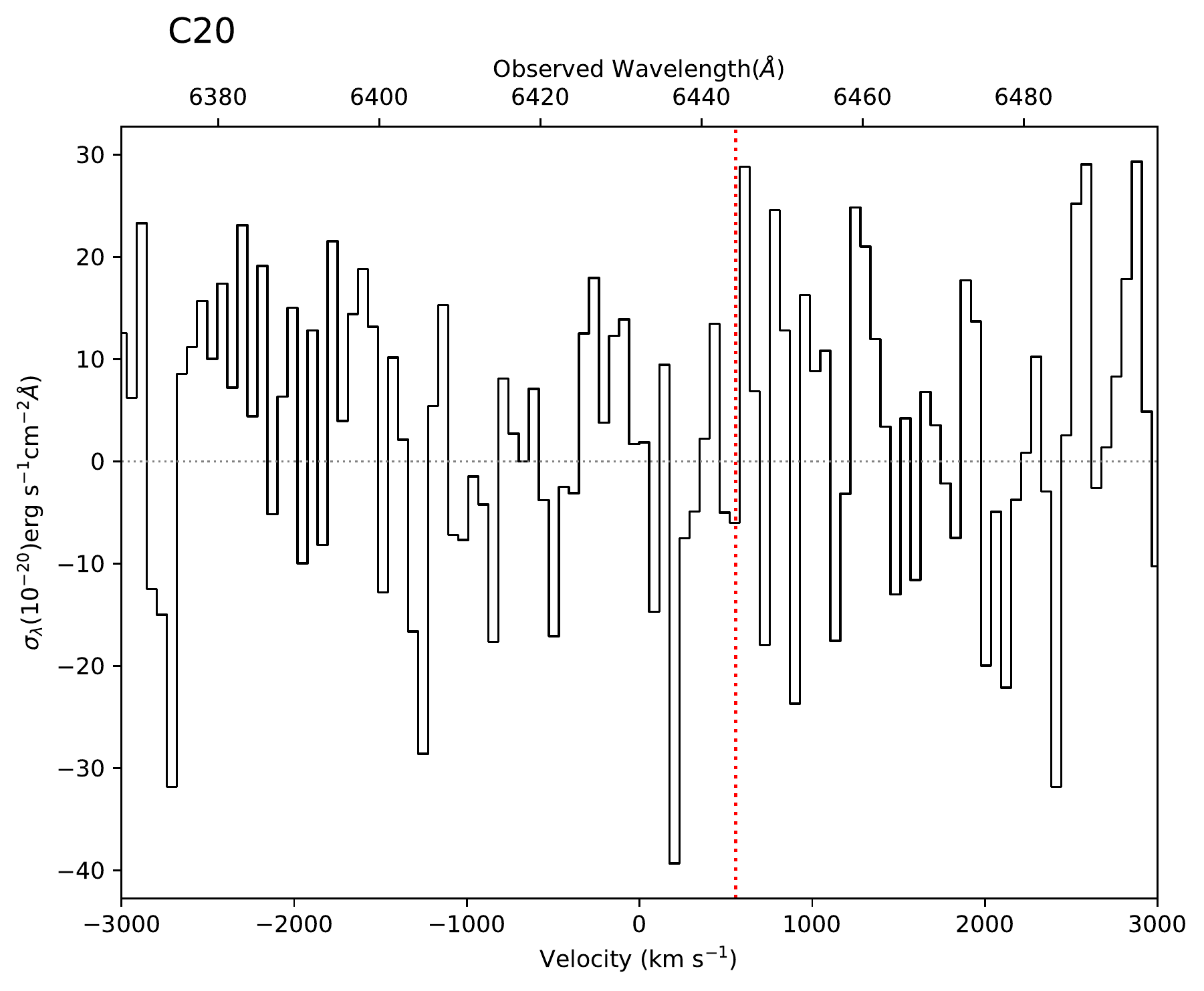} 
    \end{subfigure}
    \hfill
    \begin{subfigure}
        \centering
        \includegraphics[width=0.3\linewidth]{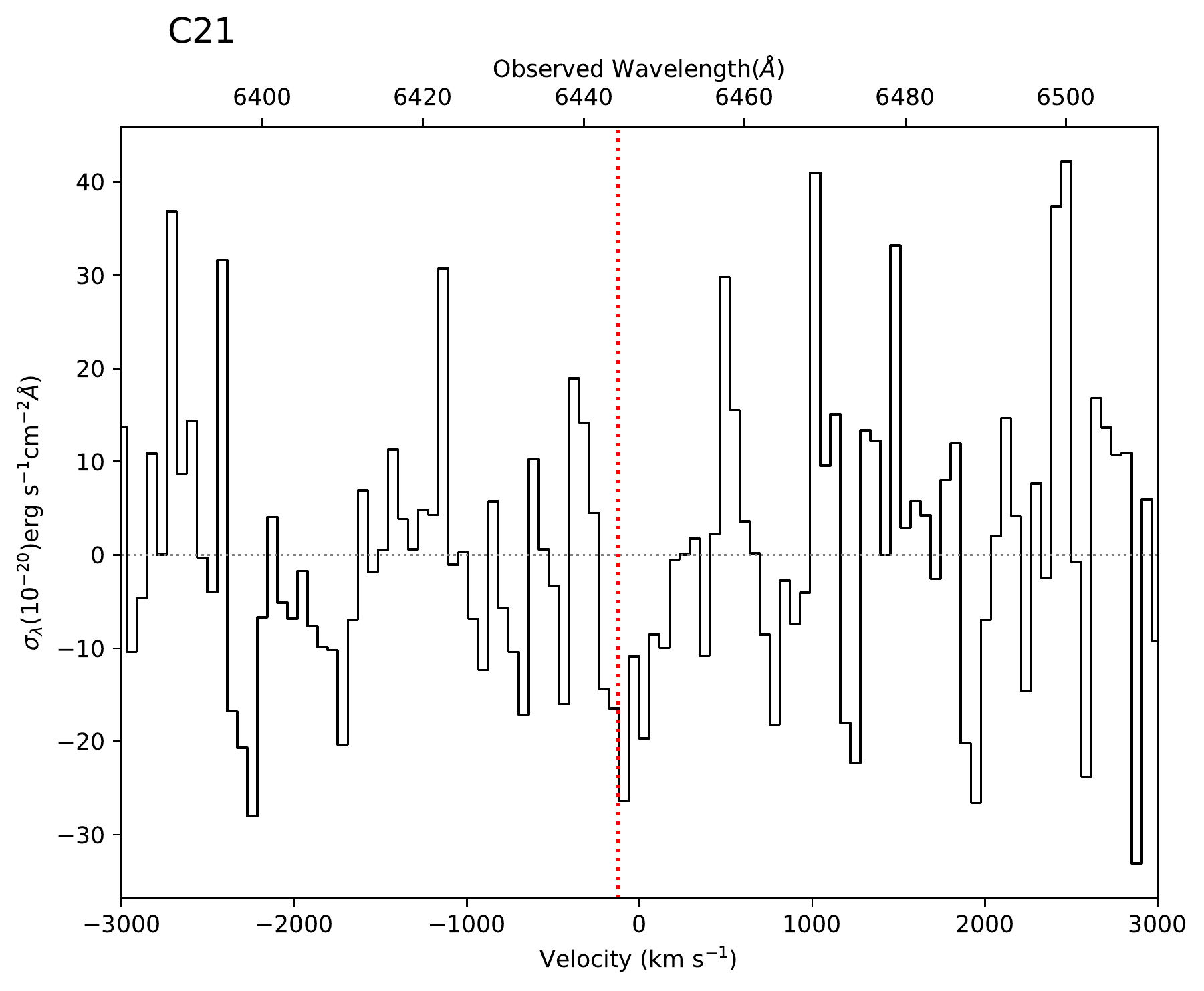} 
    \end{subfigure}
    %\vspace{1cm}
    \begin{subfigure}
        \centering
        \includegraphics[width=0.3\linewidth]{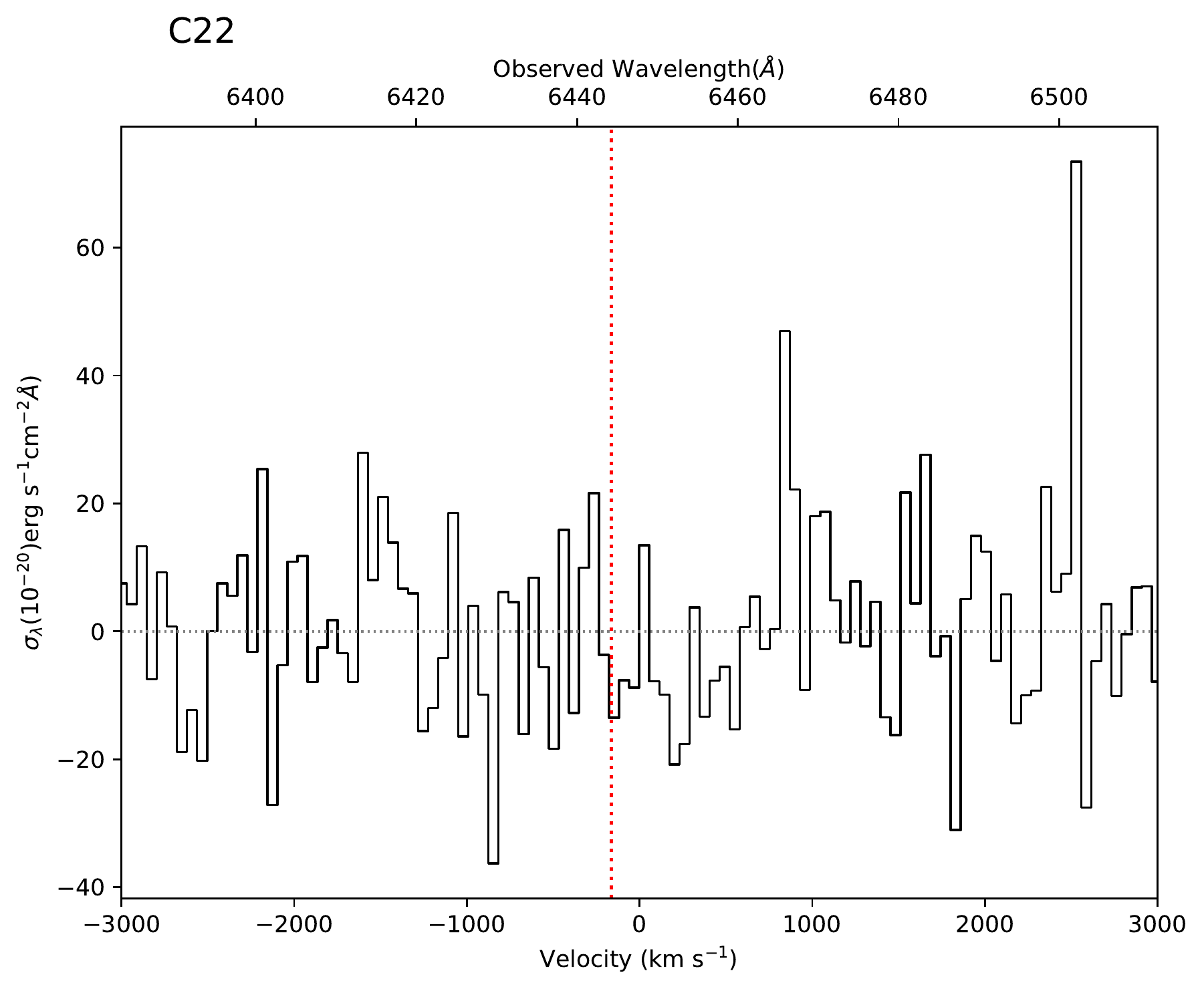} 
    \end{subfigure}
    \hfill
    \begin{subfigure}
        \centering
        \includegraphics[width=0.3\linewidth]{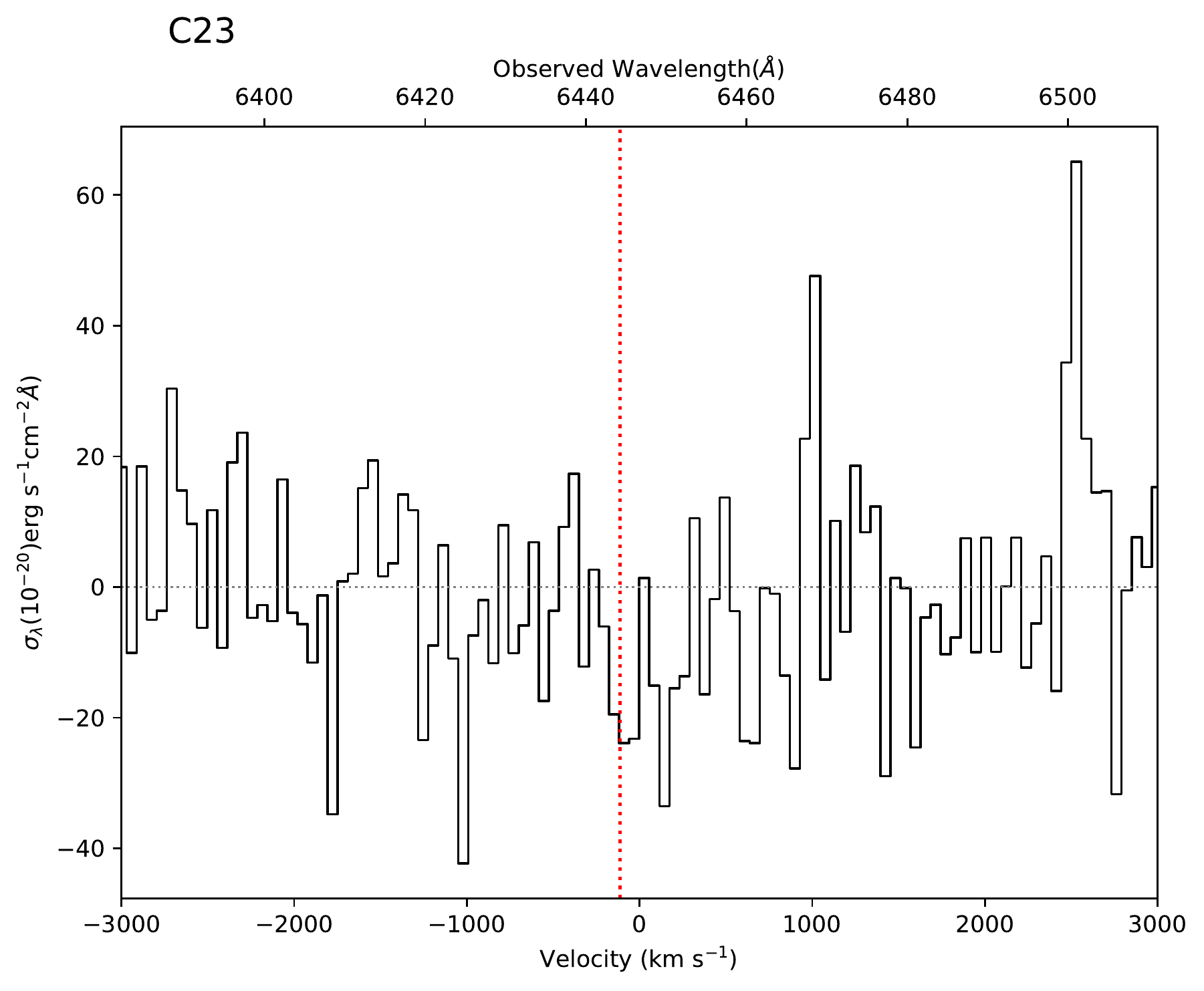} 
    \end{subfigure}
        \hfill
    \begin{subfigure}
        \centering
        \includegraphics[width=0.3\linewidth]{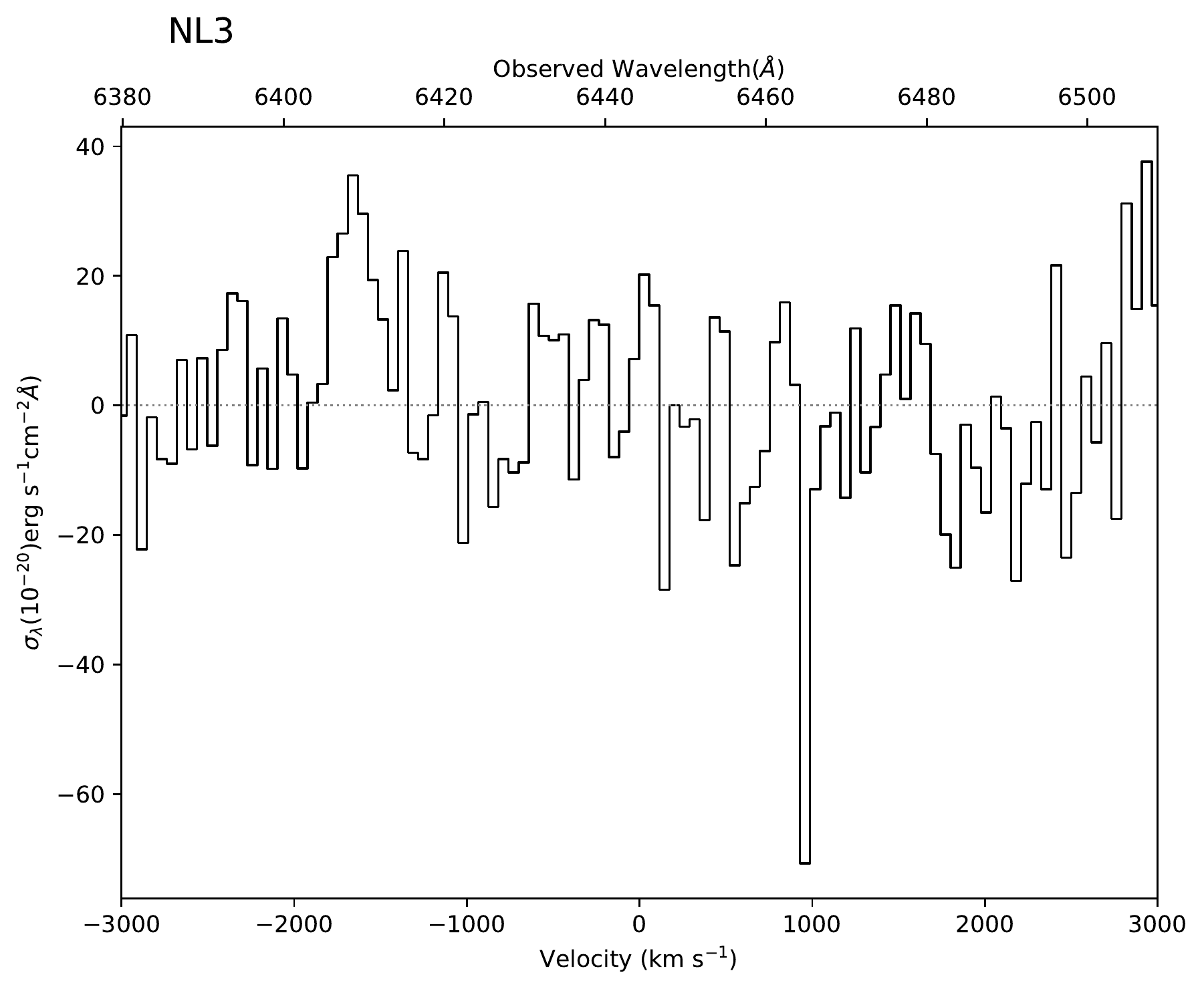} 
    \end{subfigure}
\caption{MUSE spectra of all the DSFGs previously detected toward the SPT2349-56 system at $z=4.304$. The spectra are centered at the expected wavelength for Lyman-$\alpha$ line emission. The red vertical line highlights the location of the Lyman-$\alpha$ emission line expected from the previous [C\textsc{ii}] or CO-based redshift measurement \citep{miller18, hill20}. None of the DSFGs are formally detected in Lyman-$\alpha$ emission, and only mild evidence for such line is seen in some of these spectra.} \label{fig:ALMA_detections}  
\end{figure*}
\clearpage

\section{Detected and confirmed Lyman-$\alpha$ emitters maps at different wavelength}

\begin{figure*}[ht]

\centering
    \begin{subfigure}
        \centering
        \includegraphics[width=0.7\textwidth, trim={0 0 0 12cm}]{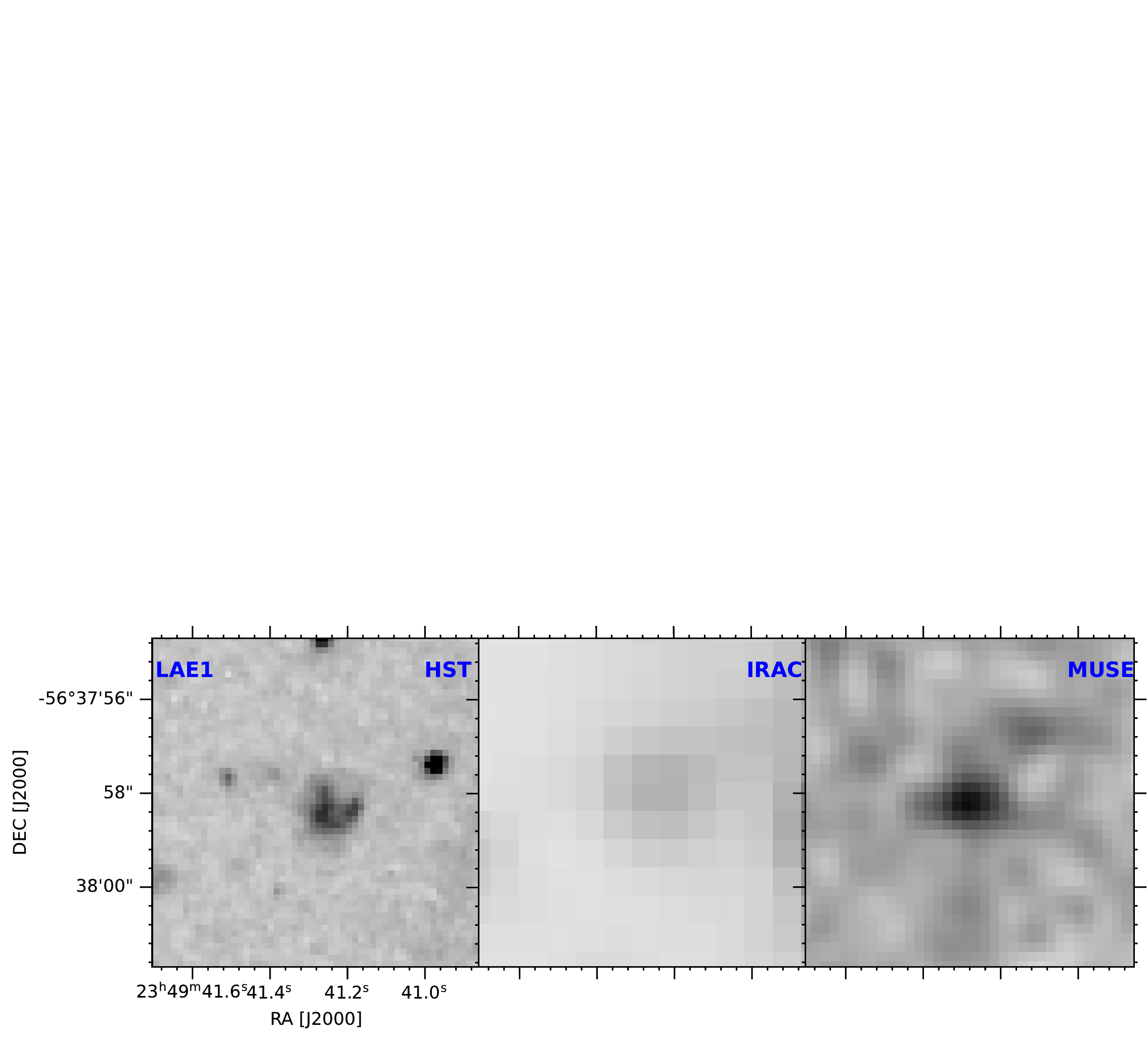}
    \end{subfigure}
    \begin{subfigure}
        \centering
    \includegraphics[width=0.7\textwidth, trim={0 0 0 13cm}]{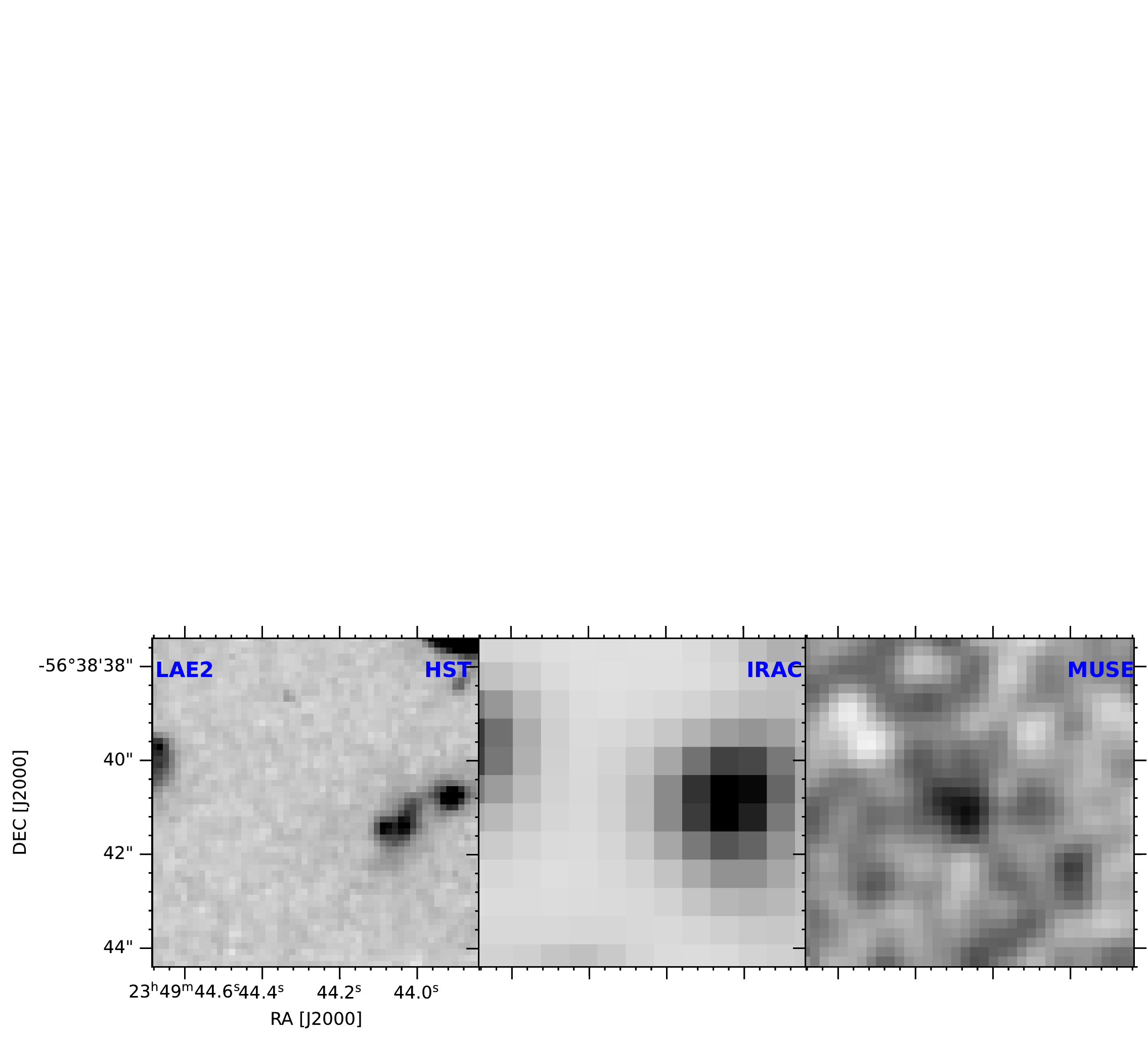}
    \end{subfigure}
    \begin{subfigure}
        \centering
    \includegraphics[width=0.7\textwidth, trim={0 0 0 13cm}]{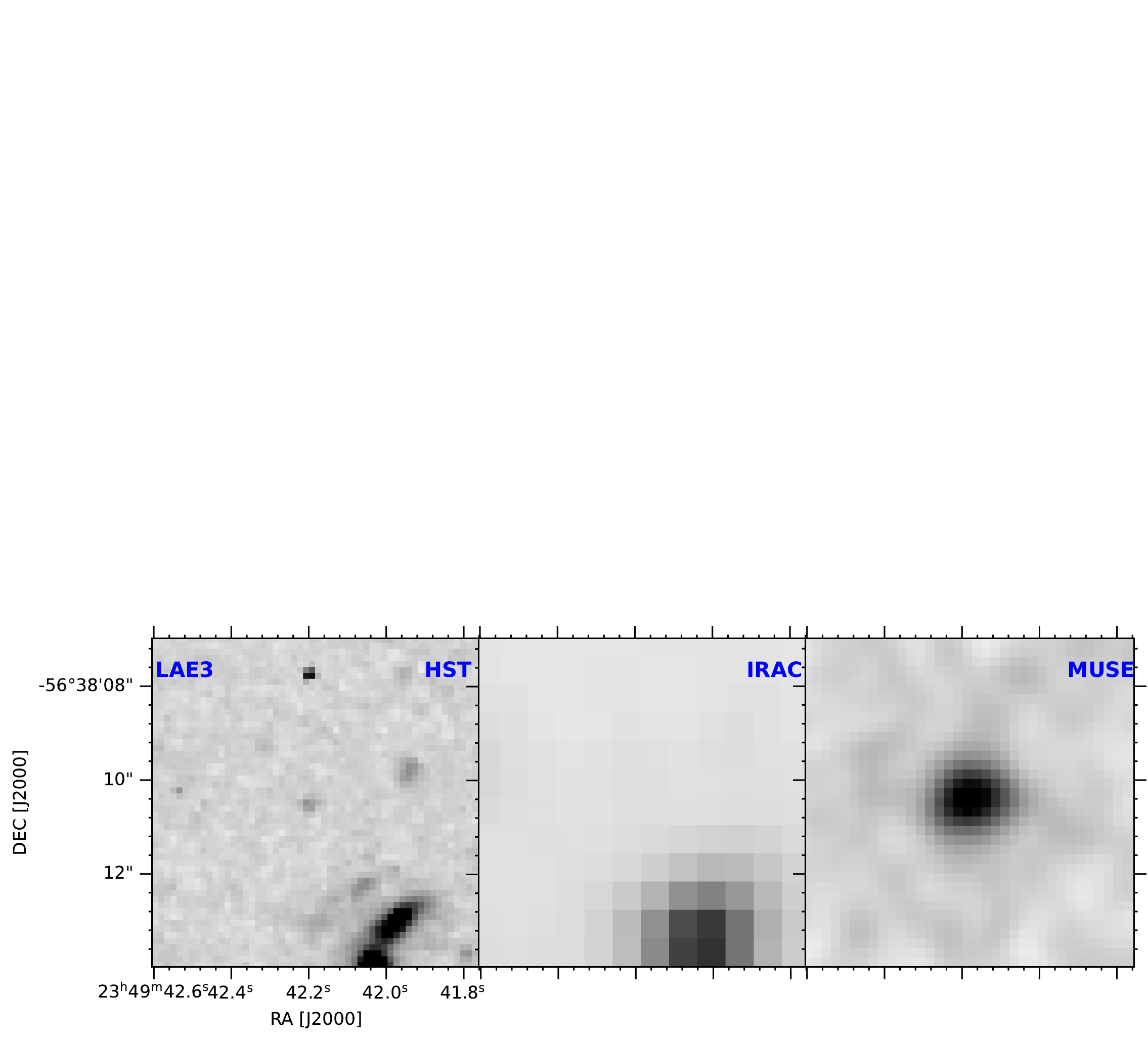}
    \end{subfigure}    
    \begin{subfigure}
        \centering
    \includegraphics[width=0.7\textwidth, trim={0 0 0 13cm}]{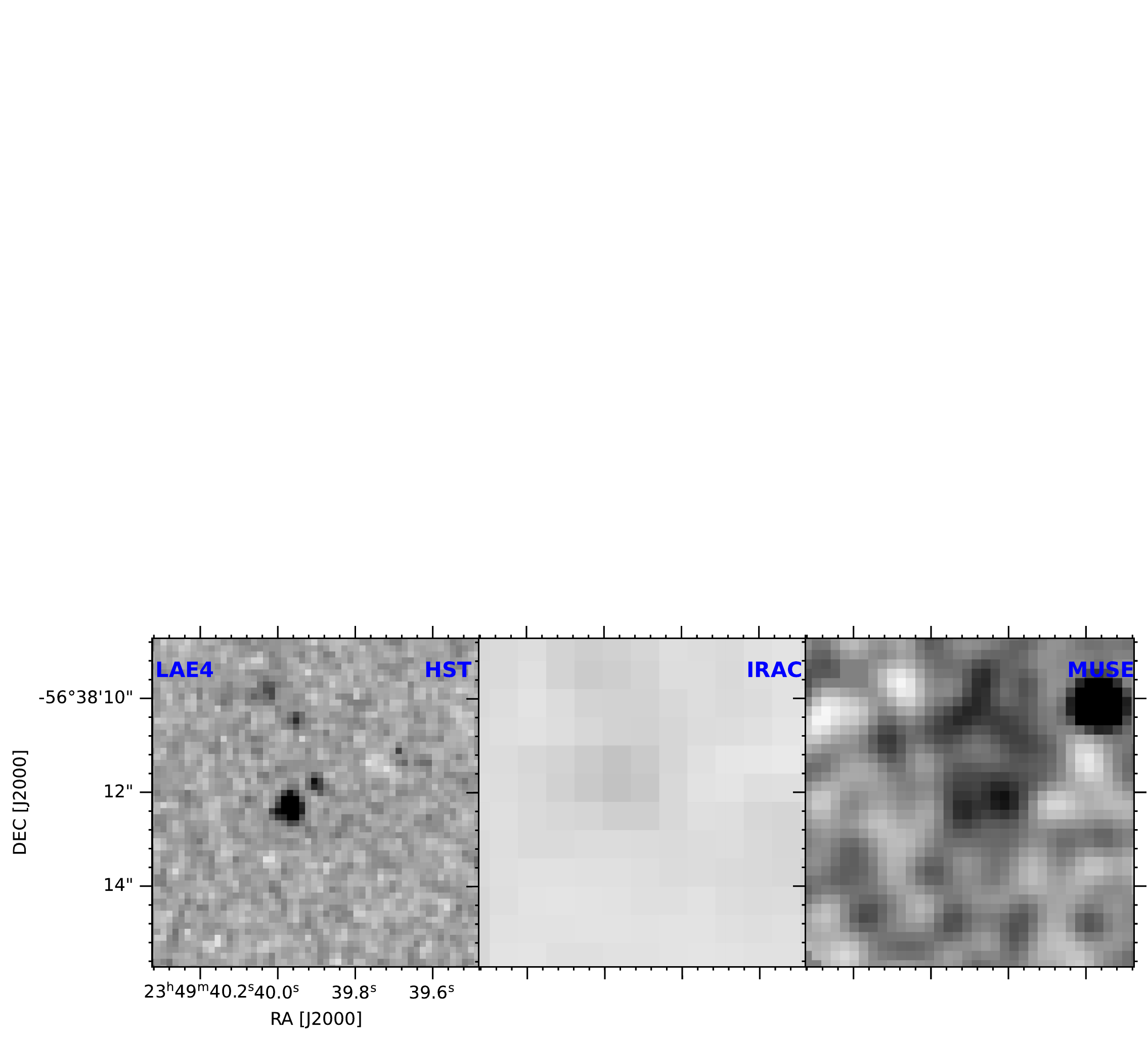}
    \end{subfigure}    

\end{figure*} 

\begin{figure*}[ht]
\centering
    \begin{subfigure}
        \centering
        \includegraphics[width=0.7\textwidth, trim={0 0 0 12cm}]{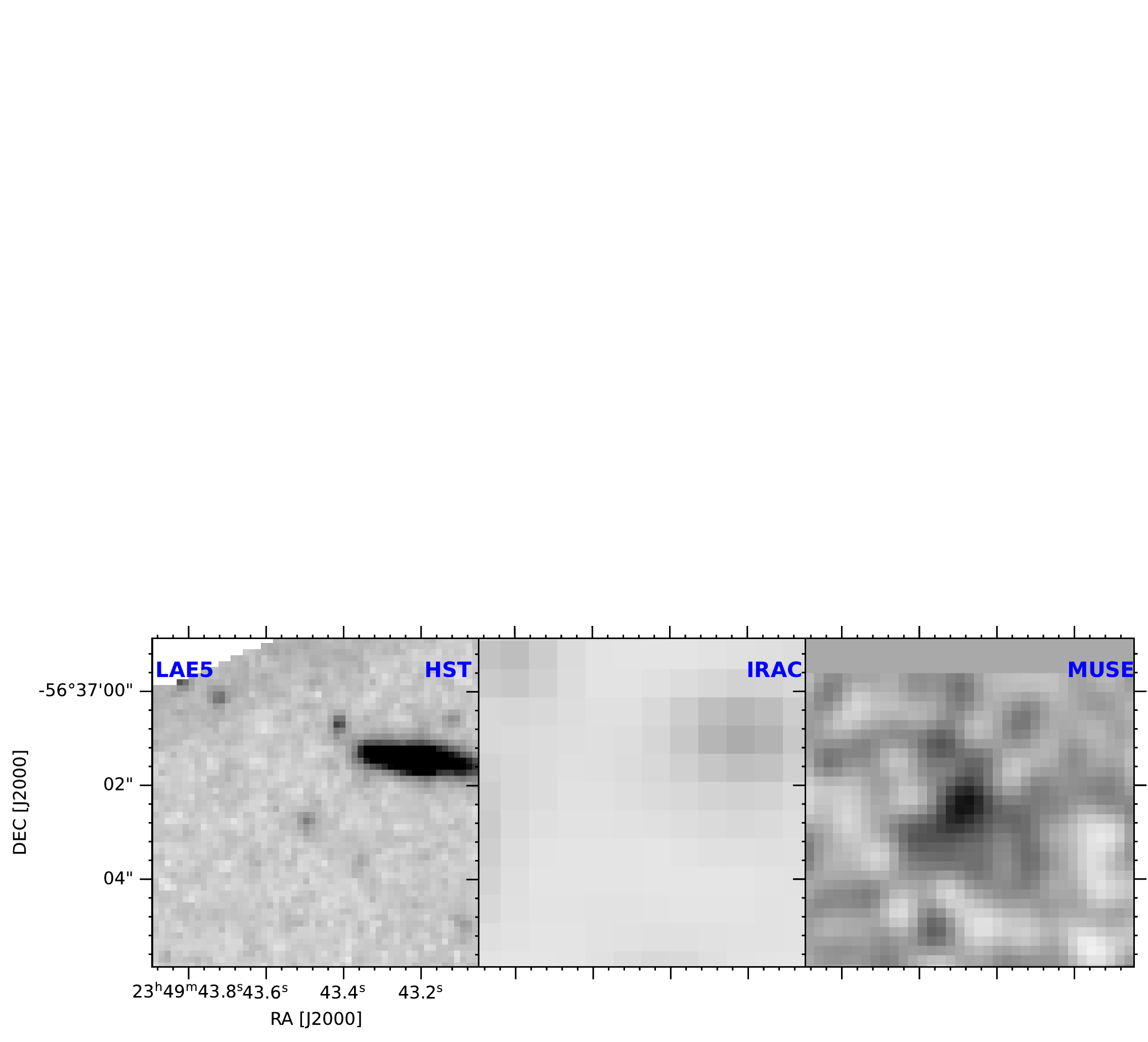}
    \end{subfigure}
    \begin{subfigure}
        \centering
    \includegraphics[width=0.7\textwidth, trim={0 0 0 13cm}]{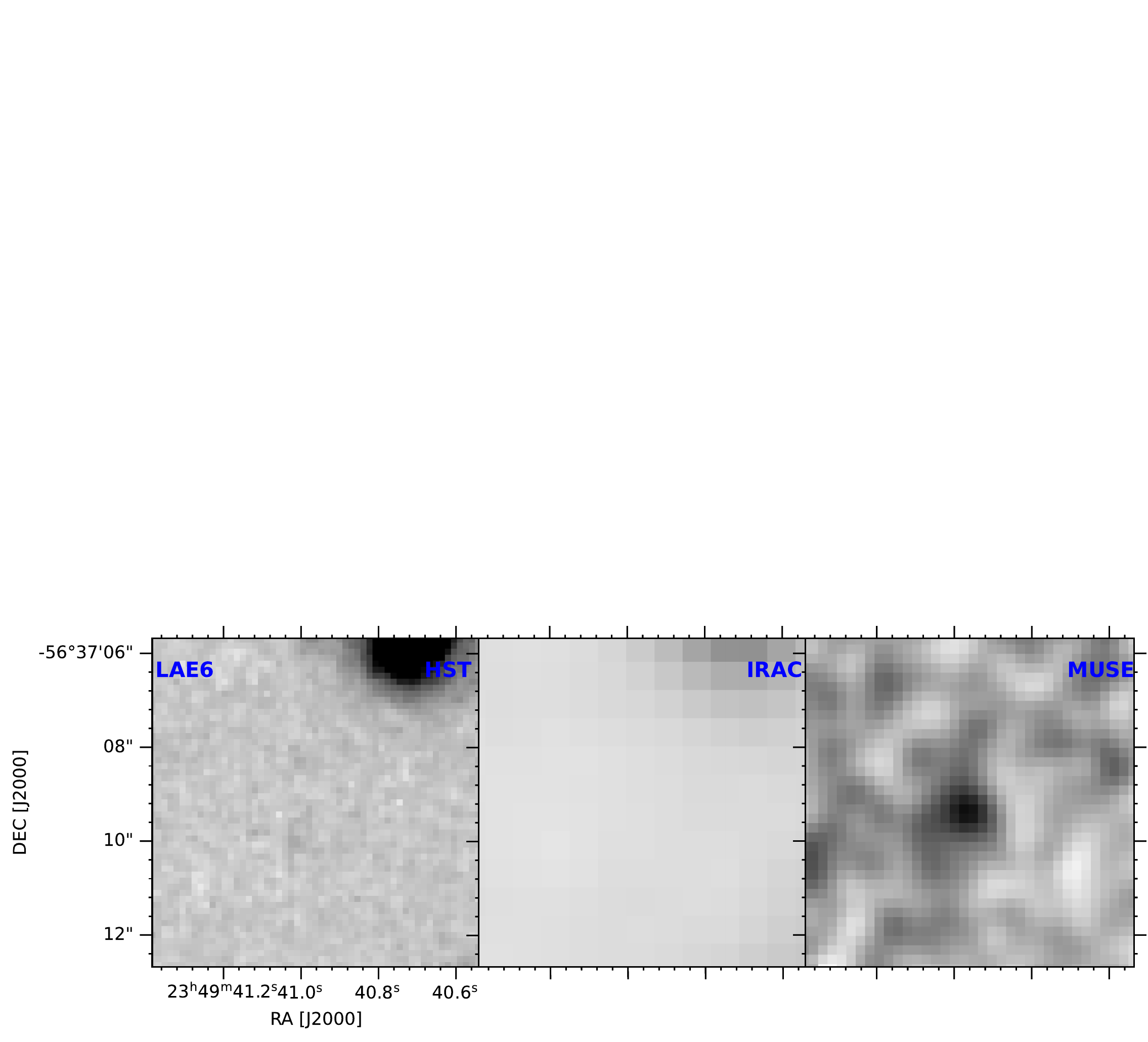}
    \end{subfigure}
    \begin{subfigure}
        \centering
    \includegraphics[width=0.7\textwidth, trim={0 0 0 13cm}]{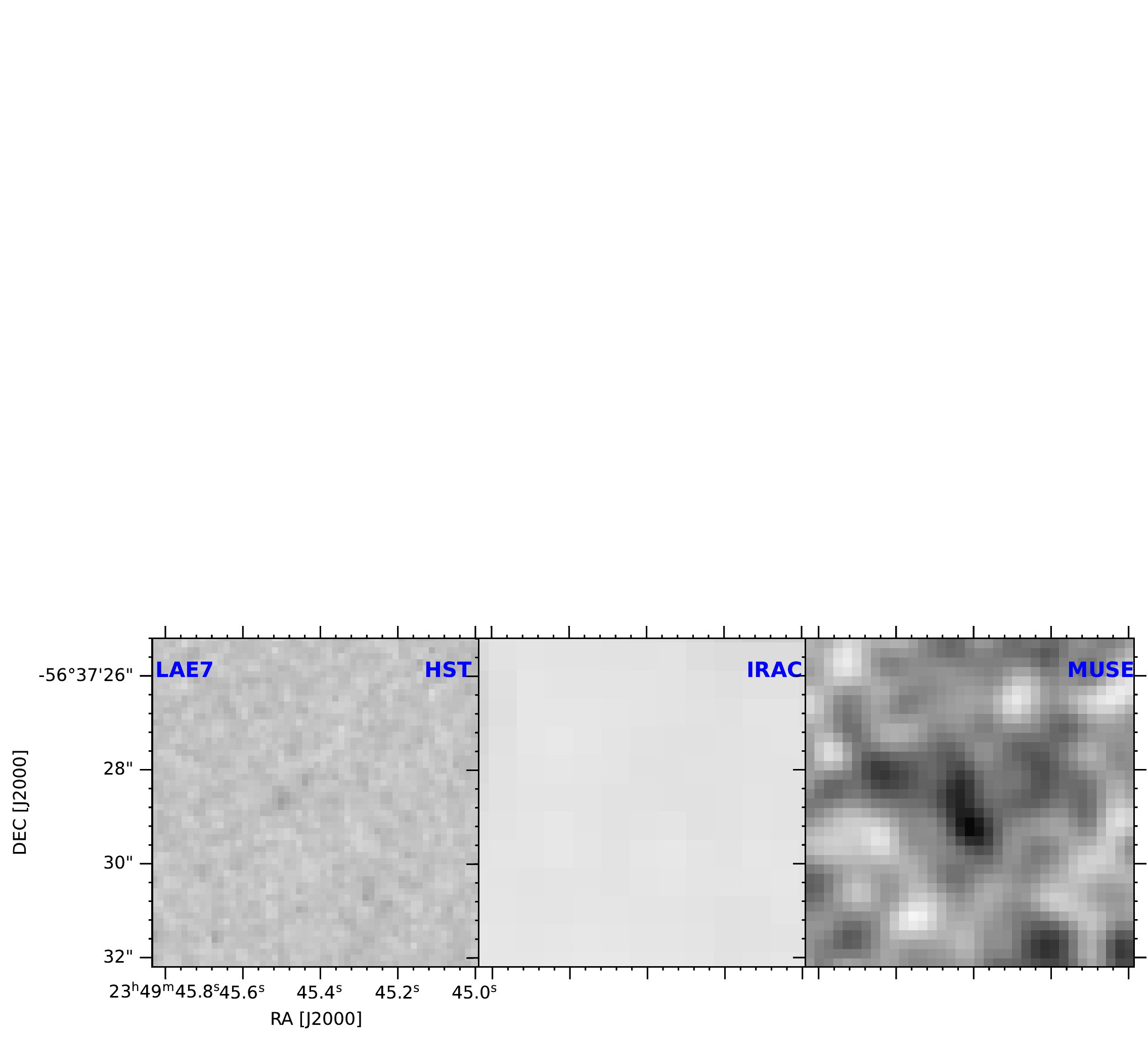}
    \end{subfigure}    
    \begin{subfigure}
        \centering
    \includegraphics[width=0.7\textwidth, trim={0 0 0 13cm}]{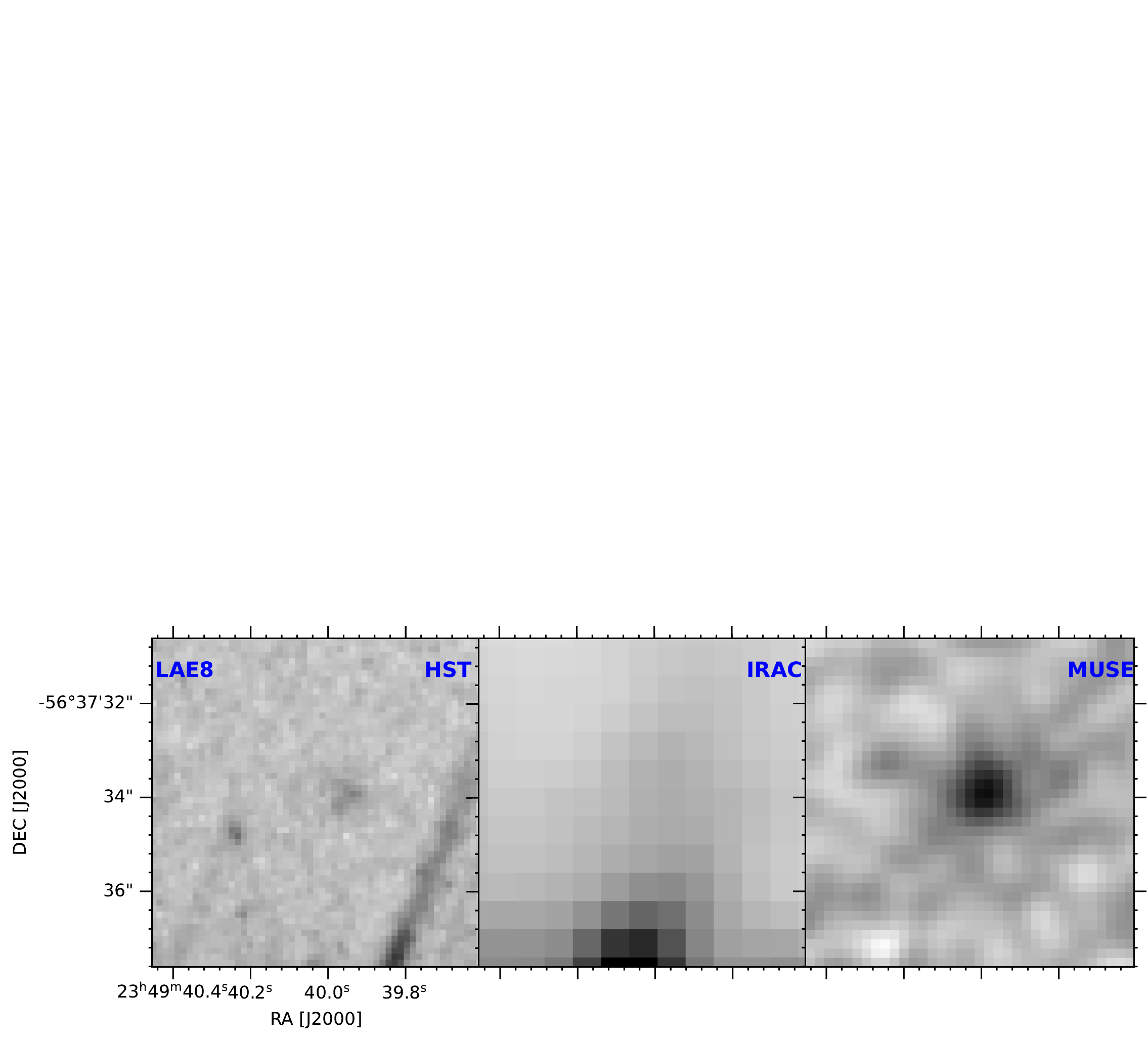}
    \end{subfigure} 
    \begin{subfigure}
        \centering
    \includegraphics[width=0.7\textwidth, trim={0 0 0 13cm}]{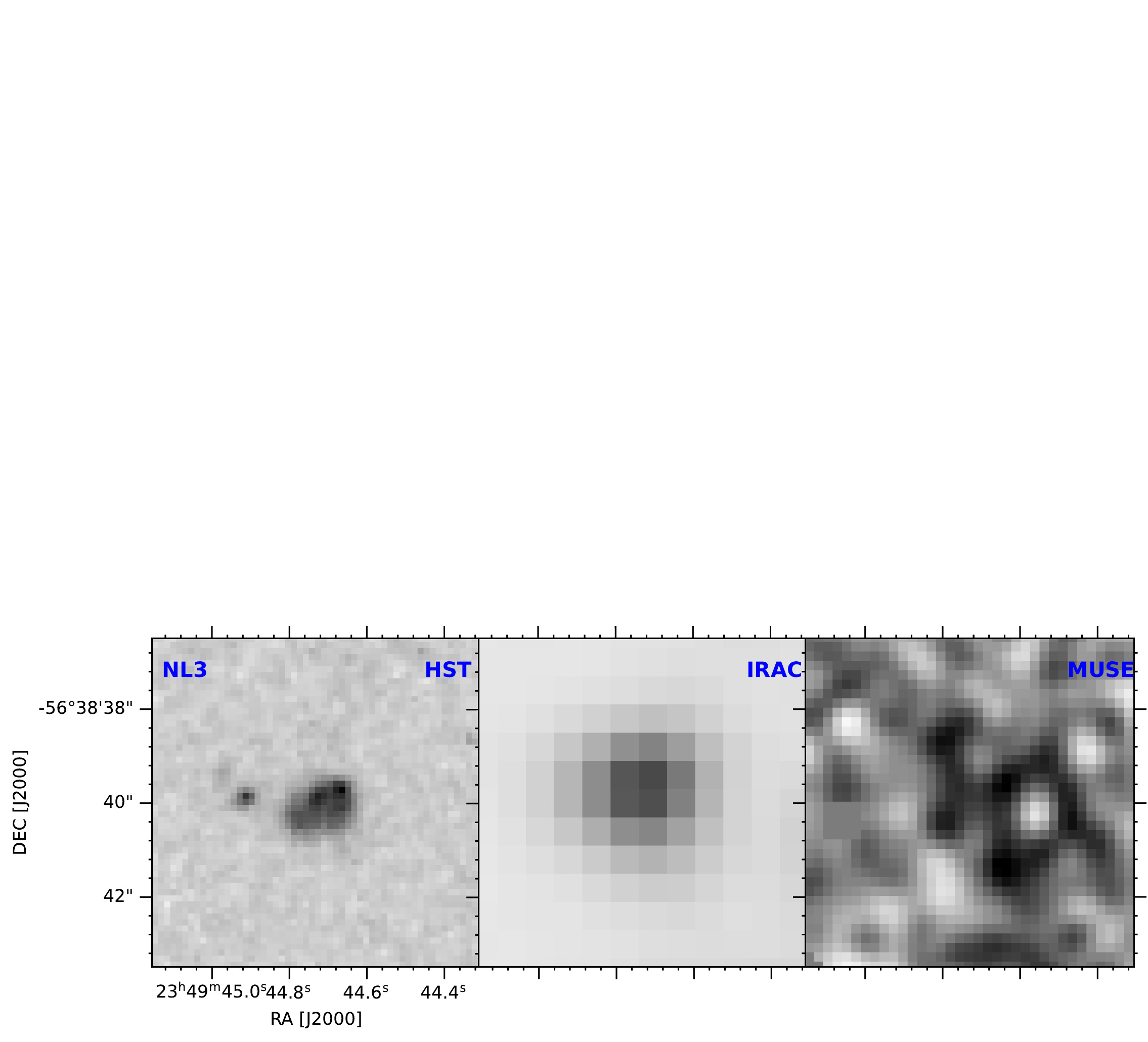}
    \end{subfigure}  
\caption{Maps centered of the detected and y tentative  Lyman-$\alpha$ emitters. \textit{Left:} HST F160W. \textit{Center:} Ultra-deep IRAC mosaic. \textit{Right:} Moment 0 of MUSE.}  
\label{fig:cutouts}
\end{figure*} 
\end{appendix}
\end{document}